\documentclass[prl,twocolumn,showpacs]{revtex4}
\usepackage{amsfonts}
\usepackage{amsmath}
\usepackage{amssymb}
\usepackage{graphicx}

\input{tcilatex}
\begin{document}

\title{Extension of the Standard Model of Electroweak interaction and Dark
Matter in the Tangent Bundle geometry\textbf{\ }}
\author{Joachim Herrmann}
\affiliation{Max Born Institute, Max Born Stra\ss e 2a , D12489 Berlin, Germany}
\email{jherrman@mbi-berlin.de}

\begin{abstract}
A generalized theory of electroweak interaction is developed based on the
underlying geometrical structure of the tangent bundle with symmetries
arising from transformations of tangent vectors along the fiber axis at a
fixed space-time point given by the SO(3,1) group. Electroweak interaction
beyond the standard model (SM) is described by the little groups $%
SU(2)\otimes E^{c}(2)$ ($E^{c}(2)$ is the central extended Euclidian group)
which includes the group $SU(2)\otimes U(1)$ as a limiting case. In addition
to isospin and hypercharge, two additional quantum numbers arise which
explain the existence of families in the SM. The connection coefficients
deliver the SM gauge potentials but also hypothetical gauge bosons and other
hypothetical particles as a Higgs family as well as candidate Dark Matter
particles are predicted. Several important consequences for the interaction
between dark fermions, dark scalars or dark vector gauge bosons with each
other and with SM Higgs and Z-bosons are described.
\end{abstract}

\maketitle

\bigskip

+\textbf{1. Introduction}

The formal equivalence of gauge theories with the geometry of fiber bundles
has been recognized since the 1960s \cite{r1,r2,r3,r5,r6}. In the fiber
bundle formalism, gauge potentials are understood as a geometrical entity -
the connections on the principal bundles, and matter fields are described by
associated fiber bundles. The geometrical interpretation of gauge theories
by the mathematical fiber bundle theory is a beautiful and mathematically
profound concept. However in earlier investigations the transformation
groups of the fibers were taken from the phenomenologically determined
internal gauge groups of the Standard Model (SM). Therefore, up to now the
fiber bundle interpretation delivers mainly a re-interpretation of the gauge
fields and did not effectuated a physical theory beyond the SM.

In this paper we consider as a general hypothesis that the fundamental
physical interactions can be described within the geometrical structure of
the most fundamental fiber bundle - the tangent bundle, and gauge
transformations can be identified with transformations at a fixed spacetime
point along the tangent vector axis leaving the scalar product invariant.
This means the gauge group is not assumed by phenomenological reasons but
arises self-consistently from the invariance of the scalar product with
respect to tangent fiber transformations described by the group $SO(3,1)$.
Since the action of this group is not transitive, the vector space
decomposes into different orbits and the most general (projective)
irreducible representations of SO(3,1) can be found by the little groups $%
SU(2),E^{c}(2)$ and $SU(1,1)$ where the group $E^{c}(2)$ is the central
extended Euclidean group$.$ Based on differential geometry on the tangent
bundle with covariant derivatives determined by the generators of the
transformation group $G=SU(2)\otimes E^{c}\mathbf{(}2)$ and corresponding
connection coefficients (gauge potentials) a generalized theory of the
electroweak interaction is derived. In addition to the internal quantum
numbers (IQN) of isospin and hypercharge,\ the $E^{c}$-charge $\varkappa $
and the family quantum number $n$ arise which could elucidate the existence
of families in the SM. In this approach the known $Z$ and $W^{\pm }$ gauge
bosons can be found again but in addition new extra $E^{c}$ and $B^{\pm }$
gauge bosons and other hypothetical particles as e.g. a family of Higgs
particles are predicted. A notable feature of the theory presented is the
possibility of identifying candidate stable or unstable hypothetical Dark
Matter (DM) vector bosons, DM scalars and DM fermions with zero hypercharge
and zero isospin but non-zero $E^{c}$-charge $\varkappa \neq 0$ without
additional phenomenological model assumptions. Here we present only the
basics, specific in-depth observable consequences are beyond the present
paper. Note that the more general transformation group $SO(3,1)\rtimes
T(3,1) $ (where $T(3,1)$ is the translational group and $\rtimes $
represents the semi-direct product) includes teleparallel gravity into the
tangent bundle geometry based on translational transformations $T(3,1)$ of
the tangent fibers. According to this approach the interpretation of the $%
SO(3,1)$ connection coefficients as electroweak gauge potentials is
compatible with teleparallel gauge gravity theory which is fully equivalent
to Einstein%
%TCIMACRO{\U{b4}}%
%BeginExpansion
\'{}%
%EndExpansion
s general relativity theory.

\textbf{2. Differential geometry on the tangent bundle}

At the beginning we start with a brief description of the geometry of the
tangent bundle on a manifold (see e.g. \cite{r7}).The tangent space $%
T_{x}(M) $ at \ the point $x$ on the space-time manifold $M$ is the set of
all tangent vectors spanned by frame vectors in the coordinate basis $e_{\mu
}=$ $\partial _{\mu }$ ($\mu =0,1,2,3)$. The tangent bundle is the union of
all tangent spaces at all points $x$ of the manifold $M:TM=\bigcup {}_{x\in
M}T_{x}(M)$. In coordinate description a point in $TM$ is described by the
numbers of pairs $X=(x,u)$ with $x=\{x^{0},x^{1},x^{2},x^{3})$ as the
coordinates of the spacetime manifold and $u=\{u^{0},u^{1},u^{2},u^{3})$ are
the coordinates of the tangent vectors. Thus the tangent fiber bundle
geometry introduces four additional variables $u$ for the description of the
tangent fiber. To aid understanding, it is convenient to consider $M$ as a
Pseudo-Riemannian spacetime manifold with indefinite metric $g^{\mu \nu }(x)$%
. Besides the frame vectors in the coordinate basis $e_{\mu }=$ $\partial
_{\mu }$ ($\mu =0,1,2,3),$ one can introduce the tetrads as another
geometric object on the tangent space: 
\begin{equation}
e_{a}=e_{a}^{\mu }(x)\partial _{\mu }.  \label{2.1}
\end{equation}%
Each vector described in the coordinate basis $e_{\mu }=$ $\partial _{\mu }$
can be expressed by a vector with respect to the tetrad frame basis $e_{a}$
according to the rule 
\begin{equation}
v^{\nu }=e_{a}^{\nu }(x)v^{a}.  \label{2.2}
\end{equation}%
The subscript $a,b,..$ numbers the vectors $(a,b=0,1,2,3)$ and $\mu $ their
components in the coordinate basis. The dual basis of the frame fields $%
e_{a} $ are cotangent frame 1-forms $e^{a}=e_{\mu }^{a}dx^{\mu }$ satisfying
the orthogonality relation $e_{a}^{\mu }(x)e_{\nu }^{a}(x)=\delta _{\nu
}^{\mu }. $ By using the tetrads of the pseudo-Riemannian manifold, the
scalar product of two vectors is given by the Lorentz metric:

\begin{align}
& (v,u)=g_{\mu \nu }(x)v^{\mu }u^{\nu }=g_{\mu \nu }(x)e_{a}^{\mu
}(x)e_{b}^{\nu }(x)v^{a}u^{b}  \notag \\
& =\eta _{ab}v^{a}u^{b},  \label{2.3}
\end{align}%
where $\eta _{ab}=diag(-1,1,1,1)$ is the metric of the Minkovski space.

The geometric properties of manifolds are usually related to the invariance
of certain geometrical structure relations under the action of certain
transformation groups. The definition of the scalar product (3) is the
governing structure relation defining the geometry of the tangent bundle.
Tangent vectors manifest two kinds of transformations which do not change
the scalar product in (3). Under general coordinate transformations of the
spacetime manifold $x^{\mu }\rightarrow y^{\mu }=y^{\mu }(x)$ vectors
transform as $v^{\prime \mu }(x)=(\partial y^{\mu }/\partial x^{\nu })v^{\nu
}(x)$. On the other hand, the vector components in the tetrad frame basis
remain unchanged: $v^{\prime a}(x)=v^{a}.$ A second type of transformations
exists that does not change the scalar product. These are transformations at
a fixed point $x$ of the spacetime manifold $M$ transforming the tangent
vectors along the tangent fiber directions as follows: 
\begin{align}
v^{\prime a}& =T_{b}^{a}(x)v^{b},e%
%TCIMACRO{\U{b4}}%
%BeginExpansion
{\acute{}}%
%EndExpansion
_{\mu }^{a}=(T_{b}^{a}(x))e_{\mu }^{b},  \label{2.4} \\
e%
%TCIMACRO{\U{b4}}%
%BeginExpansion
{\acute{}}%
%EndExpansion
_{a}^{\mu }(x)& =(T_{a}^{b}(x))^{-1}e_{b}^{\mu }(x),  \notag
\end{align}%
where $T_{b}^{a}(x)$ are matrices satisfying the conditions $\eta
_{ab}T_{c}^{a}T_{d}^{b}=\eta _{cd}.$ On the other hand, the tangent vectors
which refer to the coordinate frame remain unchanged: $v^{\prime \mu
}=v^{\mu }.$ The transformation group of tangent vectors along the tangent
fiber is the $SO(3,1)$ group of special linear transformations, with matrix
elements $T_{b}^{a}(x)\in SO(3,1)$ depending on the spacetime point $x$ as a
parameter.

Note that the transformation of the tangent vectors by the group $SO(3,1)$
is not the most general transformation. Actually, the fact that the group $%
SO(3,1)$ leaves the scalar product of tangent vectors invariant is not
sufficient because we need the infinitesimal tangent vector line elements to
be invariant. 
\begin{equation}
(dv,du)=g_{\mu \nu }(x)dv^{\mu }du^{\nu }=\eta _{ab}dv^{a}du^{b}.
\label{2.5}
\end{equation}%
This allows us to add constant translations to the transformations in (4):%
\begin{equation}
v^{\prime a}=T_{b}^{a}(x)v^{b}+a^{a}(x),  \label{2.6}
\end{equation}%
and leads to the more general transformation group $SO(3,1)\rtimes T(3,1)$.

Poincare transformations and the transformation group (6) of tangent vectors 
$T_{b}^{a}(x)$ in the tetrad basis are described by the same group $%
SO(3,1)\rtimes T(3,1)$ but both have principal different geometrical and
physical meaning: the first transforms the spacetime coordinates of a flat
manifold while the second describes transformations within the tangent fiber 
$F=T_{x}(M)$ leaving the spacetime point $x$ unchanged. The fact that
Poincare transformations (defined as coordinate transformations of spacetime
in a flat manifold) and the transformation (6) are based on the same
mathematical group could lead to confusions which can be avoided if the
principal different meaning of both transformations is taken into account.
As an example, the Coleman-Mandula theorem \cite{Col} states that the
combination of spacetime symmetries with internal symmetries is not possible
in any but trivial way. And this is what here is the case, the group $%
SO(3,1) $ in (4) or $SO(3,1)\rtimes T(3,1)$ in (6) are the internal groups
and they are combined with \ spacetime transformations as explained below
(3) only in a trivial way.

\textbf{3.} \textbf{Connections on the} \textbf{Tangent Bundle and
Teleparallel Gauge Gravity Theory}

The geometric construction of tetrads is closely linked to the conceptional
basis of gravity theories and its extensions to gravity gauge theories. To
facilitate a proper understanding of the underlying geometric structure and
a unified description including gravity we first consider the general
inhomogenous transformation group $SO(3,1)\rtimes T(3,1)$ and its
relationship with gravity. Differential geometry on the tangent bundle can
be obtained using the general rules for principal fibre bundles $P(M;G)$
requiring the definition of connections and covariant derivatives on the
bundle. The definition of a covariant derivative demands to consider vectors
which point from one fiber to the other at different points $x$ and $x%
%TCIMACRO{\U{b4}}%
%BeginExpansion
{\acute{}}%
%EndExpansion
$ of the spacetime manifold. The generators $\mathbf{L}_{a}$ of the group $G$
are vertical vectors pointing along the fiber and therefore belong to the
vertical subspace $V_{u}(P).$ Horizontal vectors in the subspace $H_{u}(P)$
which point away from the fibers (i.e. elements of the tangent space of the
fiber bundle $T_{u}(P)$ that complement the vertical vectors in $V_{u}(P)$)
can be constructed by the definition of a connection as an assigment to each
point in the principle fiber such that \cite{Ko,g4}

\begin{equation}
T_{u}(P)=H_{u}(P)\oplus V_{u}(P).  \label{3.1}
\end{equation}%
The definition of a connection can be used for the definition of a covariant
differentiation along the curves horizontally lifted to the principal
bundle: 
\begin{equation}
\frac{d}{d\tau }=\frac{dx^{\mu }}{d\tau }\emph{D}_{\mu },  \label{3.2}
\end{equation}%
where%
\begin{equation}
\emph{D}_{\mu }=\frac{\partial }{\partial x^{\mu }}+i\widetilde{A_{\mu }^{a}}%
\mathbf{L}_{a},  \label{3.3}
\end{equation}%
is the covariant derivative on the principal fiber bundle. $\mathbf{L}_{a}$
are the right-invariant fundamental vector fields (generators) on the group
manifold $G=\{g_{ij}\}$ and $\widetilde{A_{\mu }^{a}}$ the connection
coefficients of the \ group $G$. A connection on a principal bundle induces
a connection on the associated bundle. The covariant derivative on the
associated bundle is given by (9) substituting the generators $\mathbf{L}%
_{a} $ by the left-invariant fundamental vector fields on the section of the
associated bundle which describe matter fields.

The geometric transformations of tangent vectors in a tangent bundle are
described by the group $G=SO(3,1)\rtimes T(3,1).$ According to (9) the
covariant derivative along the horizontal lifted curve on the principal
bundle $P(M;G)$ of this group is given by 
\begin{equation}
D_{\mu }=\frac{\partial }{\partial x^{\mu }}+i\omega _{.\mu }^{a}\mathbf{P}%
_{a}+\frac{i}{2}\Omega _{..\mu }^{ab}\mathbf{M}_{ab},  \label{3.4}
\end{equation}%
where $\mathbf{M}_{ab}$ are related with the 6 generators of the group $%
SO(3,1)$ with $\mathbf{J}_{a}=\epsilon _{abc}\mathbf{M}_{bc},\mathbf{K}_{a}=-%
\mathbf{M}_{oa}$ and $\mathbf{P}$ $_{a}$ are the generators of the
translational group $T(3,1)$. Here $\omega _{.\mu }^{a}$ and $\Omega _{..\mu
}^{ab}$ \ are connection 1-forms of the $T(3,1)$ and $SO(3,1)$ group,
respectively. The total field strength tensor can be defined as

\begin{equation}
F_{\mu \nu }=[D_{\mu },D_{\nu }]=T_{.\mu \nu }^{a}P_{a}+\frac{1}{2}R_{..\mu
\nu }^{ab}M_{ab},  \label{3.5}
\end{equation}%
with the torsion tensor 
\begin{equation}
T_{.\nu \mu }^{a}=\partial _{\nu }\omega _{.\mu }^{a}-\partial _{\mu }\omega
_{.\nu }^{a}+(\Omega _{.e\nu }^{a}\omega _{.\mu }^{e}-\Omega _{.e\mu
}^{a}\omega _{.\nu }^{e}),  \label{3.6}
\end{equation}%
and the curvature tensor 
\begin{equation}
R_{.b\nu \mu }^{a}=\partial _{\nu }\Omega _{.b\mu }^{a}-\partial _{\mu
}\Omega _{.b\nu }^{a}+(\Omega _{.e\nu }^{a}\Omega _{.b\mu }^{e}-\Omega
_{.e\mu }^{a}\Omega _{.b\nu }^{e}).  \label{3.7}
\end{equation}%
Through contraction with tetrads, tensors can be transformed to spacetime
indexed forms as e.g. $v^{\mu }=e_{a}^{\mu }v^{a},R_{.\lambda \nu \mu
}^{\varrho }=$ $e_{a}^{\varrho }e_{\lambda }^{b}R_{.b\nu \mu }^{a}$ and the
lower frame index $v_{a}$ can be raised by the Lorentz metric $v^{a}=\eta
^{ab}v_{b}.$ The connections $\omega _{.\mu }^{a}$ and $\Omega _{..\mu
}^{ab} $ are fundamental structure functions characterizing the specific
tangent bundle.

Parallel transport of a tangent vector $v^{a}$ from a point $x$ in the
spacetime manifold to a neighboring point $x%
%TCIMACRO{\U{b4}}%
%BeginExpansion
{\acute{}}%
%EndExpansion
$ is defined by the covariant derivative

\begin{equation}
D_{\mu }v^{a}=\partial _{\mu }v^{a}+\Omega _{.\mu b}^{a}v^{b},  \label{3.8}
\end{equation}%
where $\Omega _{.\mu b}^{a}$ is denoted as frame connection. On the other
hand the covariant derivative of vectors which refer to the coordinate basis
can be written as%
\begin{equation}
D_{\mu }v^{\nu }=\partial _{\mu }v^{\nu }+\Gamma _{.\mu \rho }^{\nu }v^{\rho
},  \label{3.9}
\end{equation}%
with the coordinate connection $\Gamma _{.\mu \rho }^{\nu }.$ The coordinate
connection $\Gamma _{.\mu \lambda }^{\rho }$ is connected with the frame
connection $\Omega _{.b\mu }^{a}$ by requiring $D_{\mu }e_{\nu
}^{a}=\partial _{\mu }e_{\nu }^{a}-\Gamma _{.\mu \nu }^{\rho }e_{\varrho
}^{a}+\Omega _{.\mu b}^{a}e_{\nu }^{b}=0$ from which the following relation
can be derived%
\begin{equation}
\Omega _{.b\mu }^{a}=e_{\nu }^{a}\partial _{\mu }e_{b}^{\nu }+e_{\nu
}^{a}\Gamma _{.\rho \mu }^{\nu }e_{b}^{\varrho }.  \label{3.10}
\end{equation}%
For the description of gravity tetrads $e_{b}^{\nu }$ and a specific
coordinate connection $\Gamma _{.\mu \lambda }^{\rho }$ has to be defined.
For a manifold with vanishing metricity described by the condition $%
D_{\lambda }g_{\mu \nu }=\partial _{\lambda }g_{\mu \nu }-\Gamma _{.\mu
\lambda }^{\rho }g_{\varrho \nu }-\Gamma _{.\nu \lambda }^{\varrho }g_{\mu
\rho }=0$ one gets \cite{Sch}. 
\begin{equation}
\Gamma _{.\mu \nu }^{\rho }=\widetilde{\Gamma }_{.\mu \nu }^{\varrho
}+K_{.\mu \nu }^{\varrho },  \label{3.11}
\end{equation}%
where $\widetilde{\Gamma }_{.\mu \nu }^{\rho }$ is the Levi-Civita
connection and $K_{.\mu \nu }^{\varrho }$ is the contortion tensor%
\begin{equation}
K_{.\mu \nu }^{\varrho }=\frac{1}{2}(T_{\nu .\mu }^{.\varrho }+T_{\mu .\nu
}^{.\varrho }-T_{.\mu \nu }^{\varrho }),  \label{3.12}
\end{equation}%
with the torsion tensor $T_{.\mu \nu }^{\varrho }=\Gamma _{.\mu \nu
}^{\varrho }-\Gamma _{.\nu \mu }^{\varrho }.$The underlying geometric
structure with respect to translational and rotational transformations (6)
of tangent vectors leads to a Riemann-Cartan spacetime endowed with frame
connections $\omega _{.\nu }^{a}$ and $\Omega _{.\mu b}^{a}$, non-vanishing
curvature $R_{.b\nu \mu }^{a}$, nonvanishing torsion $T_{.\nu \mu }^{a}$ but
vanishing metricity. The special case of a Rieman spacetime and the General
Theory of Relativity (GTR) can be obtained from the above formulas by
setting the torsion tensor to be identically vanishing. The coordinate
connection $\Gamma _{.\rho \mu }^{\nu }$ then is given by the Levi-Civita
connection $\widetilde{\Gamma }_{.\mu \nu }^{\varrho }$ related with the
metric tensor $g_{\mu \nu }$. The frame connection $\Omega _{.\mu b}^{a}$ in
this case is usually denoted as spin connection $\widetilde{\Omega }_{.b\nu
}^{a}$ related with the Levi-Civita connection $\widetilde{\Gamma }_{.\mu
\nu }^{\varrho }$ by the equation (16). On the other hand based on the
symmetry with respect of translational transformations of tangent vectors a
gauge theory of gravity has been developed by analogy with internal
symmetries \cite{t1,t2,g5,t3,M}. This gauge gravity theory corresponds to
the teleparallel gravity theory, which is an alternative but equivalent
formulation of the GTR describing the very same gravitational field. In
teleparallel gravity, the torsion is non-vanishing, acting as a
gravitational force while the curvature $R_{.b\nu \mu }^{a}$ vanishes
identically. The translational connections $\omega _{\mu }^{a}$ and the
tetrad coframe $e_{\mu }^{a}$ turn out to be conceptional distinct entities,
since it does not transform inhomogeneous with a gradient term under gauge
transformation \cite{g6}. The coframes depend on the translational
connection in the form $e_{\mu }^{a}=\omega _{\mu }^{a}+D_{\mu }\xi ^{a},$ $%
D_{\mu }=\partial _{\mu }\delta _{ab}+\Omega _{.\mu b}^{a}$ where $\xi ^{a}=$
$\xi ^{a}(x)$ is a coset vector. Locally at a given point $x$ on the
spacetime manifold one can transform $D_{\mu }\xi ^{a}$ by a gauge
tansformation into $\delta _{a\mu }$ (where $\delta $ is the Kronecker
symbol). Note that the coframe $e_{\mu }^{a}$ induce \ a metric as an
independent dynamically quantity. In the geometry of teleparallel gravity
the coordinate connection $\Gamma _{.\mu \lambda }^{\rho }$ takes the form
of the Weizenb\"{o}ck connection $\Gamma _{.\mu \lambda }^{^{\parallel }\rho
}=^{W}\Gamma _{.\mu \nu }^{\rho }(x),$ defined as 
\begin{equation}
^{W}\Gamma _{.\mu \nu }^{\rho }(x)=e_{a}^{\varrho }(x)\partial _{\nu }e_{\mu
}^{a}(x).  \label{3.13}
\end{equation}%
If we consider only gravitational effects \ with the choice of the Weizenb%
\"{o}ck connection in teleparallel gravity a vector is parallel transported
if its projections on the tetrads is proportional, regardless the path
connecting both tangent spaces. This can we see from (15) with the
substitution of $\Gamma _{.\mu \nu }^{\rho }(x)$ by $^{W}\Gamma _{.\mu \nu
}^{\rho }(x)$ which yields for the covariant derivative $D_{\mu
}^{\shortparallel }v^{\nu }=e_{a}^{\nu }(x)\partial _{\mu }v^{a}.$ In
teleparallel gravity the connection coefficients $\Omega _{.b\nu
}^{\parallel a}$ represents a pure inertial effect \cite{t3} where $\Omega
_{.b\nu }^{\shortparallel a}$ is the spin connection in teleparallel
gravity. There exist a class of inertial frames in which $\Omega _{.b\nu
}^{\shortparallel a}$ vanishes: $\Omega _{.b\nu }^{\shortparallel a}=0$.

The Weizenb\"{o}ck torsion can be used to build up the Lagrangian of
teleparallel gravity with quadratic Weizenb\"{o}ck scalars\ \cite{t1,t3,M},
given by 
\begin{equation}
L=-\frac{h}{16\pi }[\frac{1}{4}T^{abc}T_{abc}+\frac{1}{2}%
T^{abc}T_{bac}-T^{a}T_{a}],  \label{3.14}
\end{equation}%
where $h=\det (e_{\mu }^{a})$ and $T_{b}=T_{ab}^{a}.$The Lagrangian is up to
a divergence equivalent to the Lagrangian of Einstein%
%TCIMACRO{\U{b4}}%
%BeginExpansion
\'{}%
%EndExpansion
s GTR, this means both theories are simply alternative formulations for the
description of gravity but are based on different principles.

A principal bundle $P(M;G)$ encodes the essential data of gauge
transformations and the frame connection $\Omega _{.b\nu }^{a}$ and $\omega
_{.\mu }^{a}$ are additional structure functions that are attached to it and
are in general independent defined on the existence of a metric ( \cite%
{Ko,g4}). If the structure group $G=$ $SO(3,1)\rtimes T(3,1)$\ is restricted
to the translational subgroup $T(3,1)$ eq..(12) and (13)\ can be taken with $%
\Omega _{.b\mu }^{a}$ put everywhere equal to zero. The torsion tensor is
now determined by $T_{.\nu \mu }^{a}=\partial _{\nu }\omega _{.\mu
}^{a}-\partial _{\mu }\omega _{.\nu }^{a}$ which can be obtained by a
specific gauge fixing and describes gravitation in the so-called pure
teleparallel gravity theory. Note \ that the structure group of the tangent
bundle is the larger group $G=$ $SO(3,1)\rtimes T(3,1)$. This raises the
question of the physical meaning of the other subgroup $SO(3,1)$. If $\Omega
_{.b\mu }^{a}$ do not vanishes we find from (13) a non-Abelian field
strength tensor which in teleparallel gauge gravity theory (based on the
translational symmetry) is not related to gravity. In this paper the main
hypothesis is elaborated that non-Abelian fields in electroweak interaction
can be identified with the connection coefficients $\Omega _{.b\mu }^{a}$
arising from transformations along the tangent fiber axis described by the
group $SO(3,1)$.This interpretation differs from Poincare gravity gauge
theory based on the localization of the Poincare group as gauge group \cite%
{g2} (for a review see \cite{g3}). In this theory both the translational
part and the rotational part of the local Poincare group is related with
gravity leading to a hypothetical generalized gravity theory, denoted as the
Einstein-Cartan gravity theory. In the following we use the denotation
connection coefficients and gauge potentials in parallel as well as gauge
transformations and tangent vector transformations.

\textbf{4. The generators on the little groups }

From now on we neglect gravity arising from the translational part $a^{a}(x)$
of the transformation of tangent vectors in (6). Since the action of $%
SO(3,1) $ on a tangent vector is not transitive, the vector space decomposes
into different orbits with the little groups $SO(3),E(2)$ and SO(1,2). The \
unitary representations $T_{L}(g)$ of the little groups $SO(3)$ and $E(2)$
are well known. The composition law of these so-called vector
representations $T_{L}(g)$ satisfy the functional equation $%
T_{L}(g_{1})T_{L}(g_{2})=T_{L}(g_{1}g_{2})$ and encodes the law of group
transformations on the set of vector states. However it is known that this
composite law is too restrictive and leads in special cases to certain
pathologies, as e.g. the Dirac equation is not invariant under the Poincare
group, but under its universal covering group. In quantum theory the
physical symmetry of a group of transformations on a set of vector states
has to preserve the transition probability between two vector states $%
\left\vert \prec \Phi ,T_{L}(g)\Psi \succ \right\vert ^{2}=$ $\left\vert
\prec \Phi ,\Psi \succ \right\vert ^{2}$. Therefore as shown by Wigner \cite%
{r8} and systematically studied by Bargman \cite{r9} the problem of
pathologies can be solved if the above given composite law is replaced by a
weaker one: $T_{L}(g_{1})T_{L}(g_{2})=\varepsilon
(g_{1},g_{2})T_{L}(g_{1}g_{2})$ where $\varepsilon (g_{1},g_{2})$ is a
complex-valued antisymmetric function of the group elements with $\left\vert
\varepsilon (g_{1},g_{2})\right\vert =1.$ Such representations are called
projective representations. For the case of simply \ connected groups like
the rotation group $SO(3$) projective representations are obtained by
replacing the group $SO(3)$ by its universal cover $SU(2)$. However in the
case of the non-semisimple Euclidean group $E(2)$ the covering group is not
enough, one has to substitute this group by a larger group: the universal
central extension $E^{c}(2)$\ which includes in addition to the group
elements of $E(2)$ the group $U(1)$ of phases factors $\varepsilon
(g_{1},g_{2})$ with$\mid \varepsilon (g_{1},g_{2})\mid =1.$ In general a
central extension $G^{c}$ of a group $G$ with elements $(g$,$\varsigma )\in $
$G^{c}$\ and $g\in $ $G,\varsigma \in $ $U(1)$ satisfy the group law \cite%
{r9} 
\begin{equation}
(g,\varsigma )=(g_{1},\varsigma _{1})\ast (g_{2},\varsigma _{2})=((g_{1}\ast
g_{2},\varsigma _{1}\varsigma _{2}\exp [i\xi (g_{1},g_{2})],  \label{4.1}
\end{equation}%
where $\xi (g_{1},g_{2})$ is the 2-cocycle satisfying the relation $\xi
(g_{1},g_{2})+\xi (g_{1}\ast g_{2},g_{3})=\xi (g_{1},g_{2}\ast g_{3})+\xi
(g_{2},g_{3}),\xi (e,g)=\xi (g,e)=0.$

The two-dimensional Euclidian group $E(2$)=$T(2$)$\otimes SO(2)$ with $E(2$)$%
=\{(\alpha ,\mathbf{a})\mid \alpha \in R\mathbf{,}(\mathrm{mod}$~$2\pi )%
\mathbf{,a=(}a\mathbf{^{1},}a\mathbf{^{2})}^{T}\mathbf{\in }R^{2}\}$ is a
semi-direct product of translations and rotations of the two-dimensional
Euclidian plane. Since the invariant subgroup $SO(2)$ is abelian this group
is not semi-simple and not compact and the unitary representations are
infinite dimensional. The representations of the group $E(2$) constructed on
the space of functions are well known. The action of the group $E(2$) on a
vector $\mathbf{Z}=(\xi _{1},\xi _{1})$ is given by:

\begin{eqnarray}
(\alpha ,\mathbf{a})(\xi _{1},\xi _{2}) &=&(\xi _{1}\cos \alpha \mathbf{-}%
\xi _{2}\sin \alpha +a^{1},  \label{4.2} \\
&&\xi _{1}\sin \alpha \mathbf{+}\xi _{2}\cos \alpha +a^{2}).  \notag
\end{eqnarray}%
The most general (projective) representations of $E(2)$ can not be obtained
from its universal covering group but by the central extended group $%
E^{c}(2).$ This group has been studied previously as e.g. in \cite%
{r9,r10,r11}. The central extension embodies in addition a U(1) subgroup
characterized by a complex parameter $\zeta =\exp (i\omega ).$ $E^{c}(2)$
consists of elements $(\alpha ,\mathbf{a,}\omega )$ with $(\alpha ,\mathbf{a}%
)\in E(2),\omega \mathbf{\in }R$. The action of the group $E^{c}(2)$ on a
vector $\mathbf{Z}=(\xi _{1},\xi _{2},\beta )$ is described by \cite%
{r9,r10,r11}.

\begin{eqnarray}
(\alpha ,\mathbf{a,\omega })(\xi _{1},\xi _{2},\beta ) &=&(\xi _{1}\cos
\alpha \mathbf{-}\xi _{2}\sin \alpha +a^{1},  \label{4.3} \\
&&\xi _{1}\sin \alpha \mathbf{+}\xi _{2}\cos \alpha +a^{2},  \notag \\
&&\beta +\omega +\frac{1}{2}m(\alpha ,\mathbf{a,}\xi _{1},\xi _{2})),  \notag
\end{eqnarray}%
where $m(\alpha ,\mathbf{a,}\xi _{1},\xi _{2})$ is the two-cocycle which
gives the desired central extension parametrized as%
\begin{eqnarray}
m(\alpha ,\mathbf{a,}\xi _{1},\xi _{2}) &=&(a_{1}\xi _{1}+a_{2}\xi _{2})\sin
\alpha -  \label{4.4} \\
&&-(a_{1}\xi _{2}-a_{2}\xi _{1})\cos \alpha .  \notag
\end{eqnarray}%
\ By using of (23) and (24) we can find the infinitesimal transformations
and the generators of the Lie algebra which satisfy the following
communication rules%
\begin{eqnarray}
\mathbf{[T}^{1},\mathbf{T}^{2}] &=&i\mathbf{E,}  \label{4.5} \\
\lbrack \mathbf{T}^{1},\mathbf{T}^{3}] &=&-i\mathbf{T}^{2},  \notag \\
\lbrack \mathbf{T}^{2},\mathbf{T}^{3}] &=&i\mathbf{T}^{1},  \notag \\
\lbrack \mathbf{T}^{a},\mathbf{E}] &=&0.  \notag
\end{eqnarray}%
\ Of particular interest are the generators on this group$:$

\begin{eqnarray}
\mathbf{T}^{1} &=&-i(\frac{\partial }{\partial \xi _{1}}+\frac{1}{2}\xi _{2}%
\frac{\partial }{\partial \beta }),\mathbf{T}^{2}=-i(\frac{\partial }{%
\partial \xi _{2}}-\xi _{1}\frac{1}{2}\frac{\partial }{\partial \beta }), 
\notag \\
\mathbf{T}^{3} &=&-i(\xi _{1}\frac{\partial }{\partial \xi _{2}}-\xi _{2}%
\frac{\partial }{\partial \xi _{1}}),\mathbf{E=-}i\frac{\partial }{\partial
\beta },  \label{4.6}
\end{eqnarray}%
The group Laplacian (Casimir operator) is determined by

\begin{equation}
\mathbf{\Delta }=(\mathbf{T}^{1})^{2}+(\mathbf{T}^{2})^{2}+2\mathbf{T}^{3}%
\mathbf{E.}  \label{4.7}
\end{equation}%
Accordingly, the operator $\mathbf{E}$ is the centre of the group. To derive
a canonical basis we use the eigenfunction of $\mathbf{E}$ ,$\mathbf{T}^{3}$
and the Laplace operator $\mathbf{\Delta }$ of $E^{c}(2)$. Using polar
coordinates $\xi _{1}=\xi \cos \phi ,\xi _{2}=\xi \sin \phi $ \ and $%
h_{nm\varkappa }(\xi ,\beta ,\phi )=\exp (i\varkappa \beta )(\exp (im\phi
)g_{nm\varkappa }(\xi )$ we find \ with the Laplacian (27) the following
equation for $g_{nm\varkappa }(\xi ):$\textbf{\ } 
\begin{eqnarray}
&&[(-\frac{1}{\xi }\frac{\partial }{\partial \xi }\xi \frac{\partial }{%
\partial \xi }+\frac{1}{\xi ^{2}}m^{2})+\varkappa ^{2}\xi ^{2}-2\varkappa
m]g_{nm\varkappa }(\xi )  \notag \\
&=&\epsilon _{nm\varkappa }g_{nm\varkappa }(\xi ).  \label{4.8}
\end{eqnarray}%
With the solutions of (28) we find for $h_{nm\varkappa }(\xi ,\beta ,\phi )$
\ 
\begin{eqnarray}
h_{nm\varkappa }(\xi ,\beta ,\phi ) &=&N_{nm\varkappa }\mid \varkappa \mid
^{\mid m\mid /2}\exp (i\varkappa \beta )(\exp (im\phi )  \notag \\
&&\exp (-\frac{\mid \varkappa \mid \xi ^{2}}{2})\xi ^{\mid m\mid
}L_{n}^{\left\vert m\right\vert }(\mid \varkappa \mid \xi ^{2}),  \label{4.9}
\end{eqnarray}%
with $N_{nm\varkappa }=\sqrt{\frac{\varkappa }{\pi }}(\frac{n!}{(\mid m\mid
+n)!})^{\frac{1}{2}}$, $\epsilon _{nm\varkappa }=4\varkappa (n+\frac{1}{2}+%
\frac{1}{2}(m+\mid m\mid ),$ with $n=0,1,2.$., $m=0,\pm 1,\pm 2,..,\varkappa
=$,$\pm 1,\pm 2,..$ Here $L_{n}^{\left\vert m\right\vert }(x)$ are the
associated Legendre polynomials. The solutions $h_{nm\varkappa }$ form an
ortho-normalized set and have the analog form like the solutions of the 2D
Schr\"{o}dinger equation in the symmetric gauge for electrons in a constant
magnetic field \cite{Q1}. The group $E^{c}(2)$ is isomorphic to the quantum
harmonic oscillator group studied e.g. in \cite{h2}.

\bigskip The special eigenvalue $\varkappa =0$ plays a particular role. The
solution of (28) for $\varkappa =0$ is given by 
\begin{equation}
h_{m00}^{\rho }=\frac{1}{\sqrt{2\pi }}J_{\left\vert m\right\vert }(\rho \xi
)\exp (im\phi ),  \label{4.10}
\end{equation}%
and agrees with the results for the group $E(2)$. $J_{\left\vert
m\right\vert }(x)$ are the Bessel functions. The eigensolutions fall into
two classes: the continuos series with $\rho ^{2}\neq 0$ and the discrete
series with $\rho ^{2}=0$ and 
\begin{equation}
h_{0m0}^{0}(\phi )=\frac{1}{\sqrt{2\pi }}\exp (im\phi ).  \label{4.11}
\end{equation}%
As described later the generator $\mathbf{T}_{3}$ corresponds to the
hypercharge generator in the SM. It is convenient to introduce for $\mathbf{T%
}_{1\text{ }}$ and $\mathbf{T}_{2}$ the generators $\mathbf{T}^{\pm }=\frac{1%
}{\sqrt{2}}(\mathbf{T}^{1}\pm i\mathbf{T}^{2})$. The action of the operator $%
\mathbf{T}^{+}$ on the eigenfunctions $g_{nm\varkappa }$ increases the
hypercharge from $m$ to $m+1$, but simultaneously generates states with
different family numbers $n$: $\mathbf{T}^{+}g_{nm\varkappa }=\Sigma
_{i=0}^{n}A_{i}g_{i,m+1,\varkappa }.$ Correspondingly the generator $\mathbf{%
T}^{-}$ reduce the hypercharge: $\mathbf{T}^{-}g_{nm\varkappa }=\Sigma
_{i=0}^{n}B_{i}g_{i,m-1,\varkappa }.$ $A_{i}$ and $B_{i}$ are coefficients,
respectively. The action of the operator product $(\mathbf{T}_{-}\mathbf{T}%
_{+}+\mathbf{T}_{+}\mathbf{T}_{-})$ on the eigenfunctions is given by%
\begin{equation}
(\mathbf{T}_{-}\mathbf{T}_{+}+\mathbf{T}_{+}\mathbf{T}_{-})g_{nm\varkappa
}(\xi ,\phi ,\beta )=q_{B}g_{nm\varkappa }(\xi ,\phi ,\beta ),  \label{4.12}
\end{equation}%
with $q_{B}(n,m,\varkappa )=4\varkappa \lbrack n+\frac{1}{2}(1+\mid m\mid
)]. $

\textbf{\ }The generators $\mathbf{J}_{1},\mathbf{J}_{2},\mathbf{J}_{3\text{ 
}}$on the group $SU(2)$ \ are well-known and, by using the Laplacian $\Delta
=(\mathbf{J}_{1}^{2}\mathbf{+J}_{2}^{2}\mathbf{+J}_{3}^{2})$ on this group,
all finite-dimensional representations can be found (see e.g. \cite{f1,f2}.
The application of the operators $\mathbf{J}_{\pm }=2^{-1/2}\mathbf{(J}%
^{1}\pm i\mathbf{J}^{2})$ and $\mathbf{J}_{3}$ on the eigenfunctions of the
Laplacian $\Delta $ leads to 
\begin{equation}
\mathbf{J}_{\pm }f_{.j_{3}}^{j}=\frac{1}{2}[j(j+1)-j_{3}(j_{3}\pm
1]^{1/2}f_{j_{3}\pm 1}^{j},  \label{4.13}
\end{equation}%
\begin{equation}
\mathbf{J}_{3}f_{j_{3}}^{j}=j^{3}f_{j_{3}}^{j},  \label{4.14}
\end{equation}%
where $j=(-j_{3}.-j_{3}+1,..,j_{3})$ are the isospin quantum numbers, $j_{3}$
is the projection on the third isospin axis. Besides we find 
\begin{equation}
(\mathbf{J}_{+}\mathbf{J}_{-}+\mathbf{J}_{-}\mathbf{J}%
_{+})f_{.j_{3}}^{j}=q_{W}f_{.j_{3}}^{j},  \label{4.15}
\end{equation}%
with $q_{W}=[j(j+1)-j_{3}^{2}].$ Using the parametrization $z_{1}=\cos \frac{%
\theta }{2}\exp [i(\psi -\varphi )/2],z_{2}=i\sin \frac{\theta }{2}\exp
[i(\psi +\varphi )]$ we find for $j=1/2,j_{3}$ $=1/2$ from the SU(2)
Laplacian $f_{j_{3}}^{j}=z_{1\text{ }}$ and for $%
j=1/2,j_{3}=-1/2:f_{j_{3}}^{j}=z_{2\text{ }}.$ For the isospin numbers $%
j=1,j_{3}=1$ we have $f_{1}^{1}=(z_{1})^{2}/\sqrt{2},$ for $j_{3}=0$ one
finds $f_{1}^{0}=z_{1}z_{2}$ and $j_{3}=-1$ yields $f_{-1}^{1}=(z_{2})^{2}/%
\sqrt{2}.$ The eigenfunctions for the general case for the isospin $j$ are
given by\cite{f2} 
\begin{equation}
f_{jj_{3}}=(\frac{1}{(j+j_{3})!(j-j_{3})!}%
)^{1/2}z_{1}^{j+j_{3}}z_{2}^{j-j_{3}},  \label{4.16}
\end{equation}

\bigskip\ In traditional Quantum field theory the internal degrees of
freedom such as iso-spin, hypercharge or color are described by spacetime
depending multi-component fields taking into account the vertical structure
given by the internal gauge groups by the corresponding Lie-algebra
representation. In the TB description in difference to the two-component
formalism the basic objects on the group $SU(2)$ (which here is interpreted
as the iso-spin group) are functions $\Phi (x,z_{1},z_{2})$ depending both
on the coordinate $x$ of the spacetime manifold and the complex coordinates $%
z_{1}=z_{1}(\theta ,\varphi ,\psi ),z_{2}=z_{2}(\theta ,\varphi ,\psi )$ .
General functions on the group manifold can be decomposed into the form 
\begin{equation}
\Phi (x,z_{1},z_{2})=\sum_{j_{3}=0}^{2j}\phi
_{j_{3}}(x)f_{j_{3}}^{j}(z_{1},z_{2})  \label{17}
\end{equation}%
The standard description for $j_{3}=\pm \frac{1}{2}$ by two-component
functions $(\phi _{1}(x),\phi _{2}(x))^{T}$ is obtained from (37)\ by the
projection into the configuration space by using the relation 
\begin{equation}
\Phi (x,z_{1},z_{2})=(z_{1},z_{2})\binom{\phi _{1}(x)}{\phi _{2}(x)}
\label{18}
\end{equation}

For final dimensional representations of the group $SU(2)$ the above
presented description by using coordinates on the group manifold $z^{1}$ and 
$z^{2}$ is equivalent to the multi-component description in non-abelian
field theory. However for a noncompact group as the group $E^{c}(2)$ a
multi-component representation is not favorable and the tangent bundle
approach using coordinates of the tangent space is more convenient.

\textbf{5. Lagrangians on the tangent bundle }

\textbf{\ 5.1 Lagrangian of Gauge fields on the group }$SU(2)\mathbf{\otimes 
}E^{c}(2)$

Based on the above described orbit decomposition and using the group $G=$ $%
SU(2)\otimes $ $E^{c}(2)$ the covariant derivative (10) with $\omega _{\mu
}^{a}=0$ can be rewritten as: 
\begin{align}
D_{\mu }=& \frac{\partial }{\partial x^{\mu }}+ig_{1}A_{\mu }^{a}(x,u)%
\mathbf{J}_{a}+ig_{2}\frac{1}{2}B_{\mu }^{a}(x,u)\mathbf{T}_{a}+  \notag \\
& +ig_{3}C_{\mu }(x,u)\mathbf{E,}  \label{53}
\end{align}%
where $\mathbf{J}_{a}$ are the generators of the group $SU(2)$ and $\mathbf{T%
}_{a}$ and $\mathbf{E}$ are the operators (26) on the group $E^{c}(2).$ $%
A_{\mu }^{a}(x,u)$,$B_{\mu }^{a}(x,u)$ and $C_{\mu }(x,u)$ are the frame
connection coefficients (gauge potentials).The gauge field strength tensors
can be obtained from the commutators $[\mathbf{D}_{\mu },\mathbf{D}_{\nu }]$%
. Let us first consider the gauge fields for the group $E^{c}(2).$ The
following relations for the field strength tensors $B_{\mu \nu }^{a}(x,u)$
and $C_{\mu \nu }(x,u)$ of the generators $\mathbf{T}_{a\text{ }}$ and $%
\mathbf{E}$ can be derived:

\begin{eqnarray}
B_{\mu \nu }^{\pm } &=&\frac{\partial }{\partial x^{\mu }}B_{\nu }^{\pm }-%
\frac{\partial }{\partial x^{\nu }}B_{\mu }^{\pm }\pm  \notag \\
&&\pm g_{2}i(B_{\mu }^{\pm }B_{\nu }^{3}-B_{\mu }^{3}B_{\nu }^{\pm })  \notag
\\
B_{\mu \nu }^{3} &=&\frac{\partial }{\partial x^{\mu }}B_{\nu }^{3}-\frac{%
\partial }{\partial x^{\nu }}B_{\mu }^{3}  \label{5.11} \\
C_{\mu \nu } &=&\frac{\partial }{\partial x^{\mu }}C_{\nu }-\frac{\partial }{%
\partial x^{\nu }}C_{\mu }-g_{2}i(B_{\mu }^{+}B_{\nu }^{-}-B_{\mu
}^{-}B_{\nu }^{+})  \notag
\end{eqnarray}%
where $B_{\mu }^{\pm }=2^{-1/2}(B_{\mu }^{1}\pm iB_{\mu }^{2}).$

Recently the non-semisimple group $E^{c}(2)$found attention in 1+1 gravity
theory \cite{IP1}, for the construction of string background in the
Wess-Zumino-Witten (WZW) model, in 3D Chern-Simon theory \cite{IP2,IP3,IP4}
as well as in Yang Mills theory \cite{IP3}. Gauge theory for a group with
generators $\mathbf{L}_{a}$ and commutators [ $\mathbf{L}_{a},$ $\mathbf{L}%
_{b}]=if_{ab}^{c}$ $\mathbf{L}_{c}$ requires a bilinear form $g_{ab}$ which
is symmetric, invariant with respect of gauge transformations and
non-degenerate so that there exist an inverse matrix. For semi-simple groups
the invariant bilinear form is given by the Killing form and proportional to
the Kronecker symbol $g_{ab}^{K}=f_{ac}^{d}f_{bd}^{c}$ $=\delta _{ab}.$ For
the non-semi-simple group $E^{c}(2)$ the Killing form is degenerate and
given by $g_{ab}^{K}=\delta _{a3}.$ Nevertheless there exists an another
form of an invariant product for this gauge group. Let us determine the
general conditions for a non-degenerate invariant product.

The Lagrangian of the gauge fields must be a quadratic combination of the
field strength tensor $B_{\mu \nu }^{a}.$ Lorentz covariance restricts its
form to 
\begin{equation}
L_{g}=-\frac{1}{4}g_{ab}B_{\mu \nu }^{a}B^{b\mu \nu },  \label{5.1}
\end{equation}%
with the invariant symmetric metric $g_{ab}$ ($g_{ab}=g_{ba}$). Under an
infinitesimal gauge transformation the field strengths transform like $%
\delta B_{\mu \nu }^{a}=-f_{bc}^{a}B_{\mu \nu }^{b}\theta ^{c}.$ The
Lagrangian must be gauge-invariant, therefore the following condition has to
be fulfilled%
\begin{equation}
g_{ab}B_{\mu \nu }^{a}f_{de}^{b}B^{d\mu \nu }\theta ^{e}=0.  \label{5.2}
\end{equation}%
Accordingly the metric $g_{ab}$ must satisfy the following condition: 
\begin{equation}
g_{ac}f_{bd}^{c}+g_{bc}f_{ad}^{c}=0.  \label{5.3}
\end{equation}%
For the group E$^{c}(2)$ with the commutation rule (25) for the generators $%
\mathbf{L}_{a}=\mathbf{T}_{a}$ $(a=1,2,3)$ and $\mathbf{L}_{4}=\mathbf{E}$
the nonzero coefficients $f_{ab}^{c}$ are given by $%
f_{12}^{4}=-f_{21}^{4}=1,f_{23}^{1}=-f_{32}^{1}=1,f_{13}^{2}=-f_{31}^{2}=-1.$
Accordingly \ one can derive a non-degenerate symmetric invariant bilinear
form $g_{ab}=g_{ba}$ given by\cite{IP1,IP2,IP3} 
\begin{equation}
g_{ab}^{0}=%
\begin{bmatrix}
1 & 0 & 0 & 0 \\ 
0 & 1 & 0 & 0 \\ 
0 & 0 & 0 & 1 \\ 
0 & 0 & 1 & 0%
\end{bmatrix}
\label{5.4}
\end{equation}

The most general invariant quadratic form is a linear combination of the
metric $g_{ab}^{0}$ and the Killing form $g_{ab}^{K}=\delta _{a3}$ given by%
\begin{equation}
g_{ab}=%
\begin{bmatrix}
1 & 0 & 0 & 0 \\ 
0 & 1 & 0 & 0 \\ 
0 & 0 & k & 1 \\ 
0 & 0 & 1 & 0%
\end{bmatrix}%
,g^{ab}=%
\begin{bmatrix}
1 & 0 & 0 & 0 \\ 
0 & 1 & 0 & 0 \\ 
0 & 0 & 0 & 1 \\ 
0 & 0 & 1 & -k%
\end{bmatrix}
\label{5.5}
\end{equation}%
with an arbitrary parameter $k$. The metric $g_{ab}$ induces a family of
non-degenerate invariant quadratic forms of Lorentz-invariant vectors $%
v_{\mu }^{a}$: 
\begin{equation}
I=v_{\mu }^{1}v^{1\mu }+v_{\mu }^{2}v^{2\mu }+kv_{\mu }^{3}v^{3\mu }+2v_{\mu
}^{3}v^{4\mu }\mathbf{.}  \label{5.6}
\end{equation}%
With a scaling of $g_{ab}\rightarrow \frac{1}{k}\widetilde{g}_{ab}$ and $%
k\gg 1$ the invariant product is given by 
\begin{equation}
I_{0}=v_{\mu }^{3}v^{3\mu }.  \label{5.7}
\end{equation}%
The generator $\mathbf{T}^{3}$ of the the group $E^{c}(2)$ corresponds to
the hypercharge operator in the SM and the Lagrangian of the corresponding
gauge field is described by the $U(1)$ group: 
\begin{equation}
L_{g}^{E^{c}}=-\frac{1}{4}B_{\mu \nu }^{3}B^{3\mu \nu }  \label{5.9}
\end{equation}

In the general case the invariant product can be diagonalized by the
transformation $v_{\mu }^{3}=\frac{1}{\sqrt{2}}(\cos \alpha {v^{\prime }}%
_{\mu }^{3}-\sin \alpha {v^{\prime }}_{\mu }^{4}),$ $v_{\mu }^{4}=\frac{1}{%
\sqrt{2}}(\sin \alpha {v^{\prime }}_{\mu }^{3}+\cos \alpha {v^{\prime }}%
_{\mu }^{4})$ with $\tan 2\alpha =\frac{2}{k}.$Then the quadratic form is
given by

\begin{equation}
I=g_{ab}v_{\mu }^{a}v^{b\mu }=v_{\mu }^{1}v^{1\mu }+v_{\mu }^{2}v^{2\mu
}+a_{3}{v^{\prime }}_{\mu }^{3}{v^{\prime }}^{3\mu }-a_{4}{v^{\prime }}_{\mu
}^{4}{v^{\prime }}^{4\mu }  \label{5.8}
\end{equation}%
with $a_{3}=k\cos ^{2}\alpha +2\cos \alpha \sin \alpha ,a_{4}=-k\sin
^{2}\alpha +2\sin \alpha \cos \alpha .$ With the choice $a_{3}=a_{4}$ we get 
$k=0$, $\alpha =\pi /4$ and $a_{3}=a_{4}=1.$

Using (45) the gauge Lagrangian of the $E^{c}(2)$ model is given by:%
\begin{eqnarray}
L_{E^{c}} &=&-\frac{1}{4}g_{ab}B_{\mu \nu }^{a}B^{b\mu \nu }  \label{5.10} \\
&=&-\frac{k}{4}B_{\mu \nu }^{3}B^{3\mu \nu }-\frac{1}{2}(B_{\mu \nu
}^{+}B^{-\mu \nu }+B_{\mu \nu }^{3}C^{\mu \nu })  \notag
\end{eqnarray}%
The Lagrangian is invariant under the gauge transformations

\begin{eqnarray}
\delta B_{\mu }^{\pm } &=&\partial _{\mu }\theta ^{\pm }\mp g_{2}i(B_{\mu
}^{3}\theta ^{\pm }-\theta ^{3}B_{\mu }^{\pm })  \label{5.12} \\
\delta B_{\mu }^{3} &=&\partial _{\mu }\theta ^{3}  \notag \\
\delta C_{\mu } &=&\partial _{\mu }\theta ^{C}+g_{2}i(\theta ^{+}B_{\mu
}^{-}-\theta ^{-}B_{\mu }^{+})  \notag
\end{eqnarray}%
where $\theta ^{3},\theta ^{\pm \text{ }},$ $\theta ^{C\text{ }}$are
arbitrary spacetime depending functions$\mathbf{.}$

The Lagrangian (50) is not positive definite and leads to negative terms in
the Hamiltonian and to the occurrence of particles with un-physical negative
norm. This situation is analogical to the case of gauge field quantization
in covariant gauge of the SM with the gauge group \textbf{\ }$SU(2)\otimes $%
\textbf{\ }$U(1)$ where due to the form of the Lorentz metric the time-like
component $B%
%TCIMACRO{\U{b4}}%
%BeginExpansion
{\acute{}}%
%EndExpansion
_{0}^{a}$ must correspond to negative metric particles. Quantization
requires to choose a specific gauge by adding terms like ($\partial _{\mu
}B_{\mu }^{a})^{2}$ and ($\partial _{\mu }C_{\mu })^{2}$ to the Lagrangian
for a covariant gauge which breaks the gauge invariance and introduce new
un-physical fields. These so-called Fadeev -Popov ghost fields cancel the
un-physical gauge field components with negative norm.

A diagonal form of the Lagrangian (50) can be achieved by the transformations%
\begin{eqnarray}
C_{\mu } &=&(C_{\mu }^{+}-C_{\mu }^{-})  \label{5.12b} \\
B_{\mu }^{3} &=&(C_{\mu }^{+}+C_{\mu }^{-})  \notag
\end{eqnarray}%
or the corresponding field strengths%
\begin{eqnarray}
C_{\mu \nu } &=&(C_{\mu \nu }^{+}-C_{\mu \nu }^{-})  \label{5.12a} \\
B_{\mu \nu }^{3} &=&(C_{\mu \nu }^{+}+C_{\mu \nu }^{-})  \notag
\end{eqnarray}

The Lagrangian of gauge particles with a gauge fixing term and the
Fadeev-Popov ghosts is for $k=0$ given by

\begin{eqnarray}
L_{E^{c}} &=&-\frac{1}{4}\{2B_{\mu \nu }^{+}B^{-\mu \nu }+C_{\mu \nu
}^{+}C^{+\mu \nu }-C_{\mu \nu }^{-}C^{-\mu \nu }\ +  \notag \\
&&+\frac{1}{2\xi }(\partial ^{\mu }B_{\mu }^{+})^{2}+\frac{1}{2\xi }%
(\partial ^{\mu }B_{\mu }^{-})^{2}+\frac{1}{2\zeta }(\partial ^{\mu }C_{\mu
}^{+})^{2}+  \notag \\
&&+\frac{1}{2\zeta }(\partial ^{\mu }C_{\mu }^{-})^{2}-(\partial _{\mu }%
\overline{c}^{3})c^{3}-\partial _{\mu }\overline{c}^{+}c^{-}-(\partial _{\mu
}\overline{c}^{-})c^{+}+  \notag \\
&&-(\partial _{\mu }\overline{c}^{4})c^{4}+ig_{2}(\partial _{\mu }\overline{c%
}^{4})(c^{+}B^{-\mu }-c^{-}B^{+\mu })+  \notag \\
&&-ig_{2}[(\partial _{\mu }\overline{c}^{+})c^{-}-(\partial _{\mu }\overline{%
c}^{-})c^{+}](C^{+\mu }+C^{-\mu })+  \label{5.12d} \\
&&-ig_{2}[(\partial _{\mu }\overline{c}^{+})B^{-\mu }-(\partial _{\mu }%
\overline{c}^{-})B^{+\mu }]c^{3}\},  \notag
\end{eqnarray}%
where $c^{\pm }$ =$\frac{1}{\sqrt{2}}(c^{1}\pm ic^{2}),c_{3},c_{4}$ and $%
\overline{c}^{\pm },\overline{c}^{3},$ $\overline{c}^{4}$ are the
anticommuting ghost fields and $\zeta $ the gauge parameter. The
cancellation of the un-physical gauge particles with negative norm by the
Fadeev-Popov ghost particles can be proven by the BRST symmetry in a
gauge-invariant form. Quantization of non-semi-simple gauge groups has been
studied in \cite{IP3,IP5,IP6,IP7}. In \cite{IP3} one-loop radiative
corrections for the Yang Mills model with the $E^{c}(2)$ gauge group were
computed. It was shown that there is no two and higher loop re-normalization
and the full quantum effective action is given by the 1-loop term with the
divergent part that can be eliminated by a field redefinition.

The field strength tensor for the group $SU(2)$ is given by 
\begin{eqnarray}
W_{\mu \nu }^{\pm } &=&\frac{\partial }{\partial x^{\mu }}W_{\nu }^{\pm }-%
\frac{\partial }{\partial x^{\nu }}W_{\mu }^{\pm }\pm  \notag \\
&&\pm g_{1}i(W_{\mu }^{\pm }B_{\nu }^{3}-B_{\mu }^{3}W_{\nu }^{\pm }) \\
B_{\mu \nu }^{3} &=&\frac{\partial }{\partial x^{\mu }}B_{\nu }^{3}-\frac{%
\partial }{\partial x^{\nu }}B_{\mu }^{3}-g_{1}i(W_{\mu }^{+}W_{\nu
}^{-}-W_{\mu }^{-}W_{\nu }^{+})  \notag
\end{eqnarray}

The total gauge Lagrangian on the TB is $L_{g}=L_{E^{c}}+L_{SU(2)}$ where
the Lagrangian of the group $SU(2)$ with the inclusion of ghost fields is
given by

\begin{eqnarray}
L_{SU(2)} &=&-\frac{1}{4}\{2W_{\mu \nu }^{+}W^{-\mu \nu }+A_{\mu \nu
}^{3}A^{3\mu \nu }+\frac{1}{2\xi }(\partial ^{\mu }W_{\mu }^{+})^{2}+  \notag
\\
&&+\frac{1}{2\xi }(\partial ^{\mu }W_{\mu }^{-})^{2}+\frac{1}{2\zeta }%
(\partial ^{\mu }A_{\mu }^{3})^{2}+  \notag \\
&&-(\partial _{\mu }\overline{\omega }^{3})\omega ^{3}-\partial _{\mu }%
\overline{\omega }^{+}\omega ^{-}-(\partial _{\mu }\overline{\omega }%
^{-})\omega ^{+} \\
&&+ig_{2}(\partial _{\mu }\overline{\omega }^{3})(\omega ^{+}W^{-\mu
}-\omega ^{-}W^{+\mu })  \notag \\
&&-ig_{2}[(\partial _{\mu }\overline{\omega }^{+})\omega ^{-}-(\partial
_{\mu }\overline{\omega }^{-})\omega ^{+}]B^{3\mu }  \notag \\
&&-ig_{2}[(\partial _{\mu }\overline{\omega }^{+})W^{-\mu }-(\partial _{\mu }%
\overline{\omega }^{-})W^{+\mu }]\omega ^{3}\}  \notag
\end{eqnarray}%
with field strength tensors of the $SU(2)$ gauge field $W_{\mu }^{\pm
}=2^{-1/2}(A_{\mu }^{1}\pm iA_{\mu }^{2})$ and $A_{\mu }^{3}$. $\overline{%
\omega }^{a}$ and $\omega ^{a}$ are a set of independent anticommuting
variables of the ghosts.

\textbf{5.2 Lagrangians of matter fields on the group }$SU(2)\mathbf{\otimes 
}E^{c}(2)$

In the Lagrangian of the Higgs scalar particles $L_{H}=(D_{\mu }\Phi
_{H})^{\dagger }(D^{\mu }\Phi _{H})$ the electromagnetic field should not
couple to the neutrino and should be diagonalized. Substituting (39) into
the Lagrangian of the Higgs new mixing terms appear and the diagonalization
requires that the fields $A_{\mu }^{3},$ $B_{\mu }^{3}$ and $C_{\mu }$ \
have to be transformed to new fields expressed by the relations 
\begin{eqnarray}
B_{\mu }^{3} &=&\cos \theta _{W}A_{\mu }-\sin \theta _{W}(\cos \theta
_{D}Z_{\mu }-\sin \theta _{D}E_{\mu }^{c}),  \notag \\
A_{\mu }^{3} &=&\sin \theta _{W}A_{\mu }+\cos \theta _{W}(\cos \theta
_{D}Z_{\mu }-\sin \theta _{D}E_{\mu }^{c}),  \notag \\
C_{\mu }^{3} &=&\sin \theta _{D}Z_{\mu }+\cos \theta _{D}E_{\mu }^{c}.
\label{B1}
\end{eqnarray}%
where $\theta _{W}$ is the Weinberg angle, $g_{2}=g\sin \theta
_{W},g_{1}=g\cos \theta _{W},$ $g=((g_{1})^{2}+(g_{2})^{2})^{1/2}$,$e=g\cos
\theta _{W}\sin \theta _{W}$ and $\theta _{D}$ is defined by $\tan 2\theta
_{D}=gg_{3}m_{H}\varkappa _{H}/[(g_{3}\varkappa _{H})^{2}-(\frac{m_{H}}{2}%
g)^{2}]$, $m_{H}$ and $\varkappa _{H}$ are the IQNs of the Higgs particle.
The covariant derivative (39) can be rewritten as 
\begin{eqnarray}
\mathbf{D}_{\mu } &=&\frac{\partial }{\partial x^{\mu }}+i[W_{\mu }^{+}%
\mathbf{Q}_{W^{-}}+W_{\mu }^{-}\mathbf{Q}_{W^{+}})+B_{\mu }^{+}\mathbf{Q}%
_{B^{-}}  \label{5.17} \\
&&+B_{\mu }^{-}\mathbf{Q}_{B^{+}}+Z_{\mu }\mathbf{Q}_{Z}+A_{\mu }\mathbf{Q+}%
E_{\mu }^{c}\mathbf{Q}_{E^{c}}],  \notag
\end{eqnarray}%
with $W_{\mu }^{\pm }=2^{-1/2}(W_{\mu }^{1}\pm iW_{\mu }^{2})$ and $B_{\mu
}^{\pm }=2^{-1/2}(B_{\mu }^{1}\pm iB_{\mu }^{2}).$The following operators
are introduced 
\begin{align}
\mathbf{Q}& =(\mathbf{J}_{3}+\frac{1}{2}\mathbf{T}_{3}),  \label{5.18} \\
\mathbf{Q}_{Z}& =g\cos \theta _{D}(\cos ^{2}\theta _{W}\mathbf{J}_{3}-\sin
^{2}\theta _{W}\frac{1}{2}\mathbf{T}_{3})  \notag \\
& +g_{3}\sin \theta _{D}\mathbf{E,}  \notag \\
\mathbf{Q}_{W^{\pm }}& =g_{1}\mathbf{J}^{\pm },\mathbf{Q}_{B^{\pm }}=g_{2}%
\frac{1}{2}\mathbf{T}^{\pm },  \notag \\
\ \mathbf{Q}_{E^{c}}& =g(-\cos ^{2}\theta _{W}\mathbf{J}_{3}+\sin ^{2}\theta
_{W}\frac{1}{2}\mathbf{T}_{3})\sin \theta _{D}  \notag \\
& +g_{3}\cos \theta _{D}\mathbf{E}.  \notag
\end{align}

Any scalar function $\Phi (x,u)$ defined on the fiber bundle can be expanded
into the form $\Phi =\sum_{M}(\phi _{M}(x)\chi _{M}(u)+\phi _{M}^{\dagger
}(x)\chi _{M}^{\ast }(u))$ depending on the coordinates of the spacetime
manifold $x$ and the eigenfunctions $\chi _{M}(u)$ $=h_{lm\varkappa }(\xi
,\phi ,\beta )f_{jj_{3}}(z_{1},z_{2})$ of the Laplacian of the group $%
SU(2)\otimes $\textbf{\ }$E^{c}(2)$ defined by (29) and (36)$.$ Besides we
introduced the symbol for the IQNs: $M=(n,m$,$\varkappa ,j,j_{3}).$ The
total Lagrangian can be presented by 
\begin{equation}
L=L_{l}+L_{H}+L_{g}+L_{Yuk}.  \label{5.24}
\end{equation}%
Here the Lagrangian of leptons $L_{l}$ is defined in the chiral
representation as 
\begin{align}
L_{l}& =i\dsum\limits_{s}\Psi _{s}^{l\dagger }(x,u)\sigma _{s}^{\mu }\mathbf{%
\{}\frac{\partial }{\partial x^{\mu }}+i[W_{\mu }^{+}\mathbf{Q}%
_{W^{-}}+W_{\mu }^{-}\mathbf{Q}_{W^{+}}  \notag \\
& +Z_{\mu }\mathbf{Q}_{Z}+A_{\mu }\mathbf{Q+}E_{\mu }^{c}\mathbf{Q}%
_{E^{c}}]\}\Psi _{s}^{l}(x,u),  \label{25}
\end{align}%
with the helicity $s=\{L,R\}$ and $\sigma _{R}^{\mu }=(\sigma ^{0},\sigma
^{i}),\sigma _{L}^{\mu }=(\sigma ^{0},-\sigma ^{i})$ and $\Psi
_{s}^{l}(x,u)=\sum_{M}(\psi _{M,s}^{-}(x)\chi _{M}(u)+\psi _{M,s}^{+}(x)\chi
_{M}^{\ast }(u)).$ Note that gauge fields and fermions or scalars can carry
different IQNs. As later is shown the known SM gauge particles as the the $Z$
and $W^{\pm }$ bosons as well as the $E^{c}$ gauge boson carry the IQN $%
\varkappa =0,$ but due to the existence of families SM leptons carry
non-zero E$^{c}-$charges $\varkappa \neq 0.$ The interaction of leptons with
the $B^{\pm }$ bosons is forbidden because of a selection rule discussed
below.

The Lagrangian of the SM Higgs particles $\Phi _{H}=$ $\Phi _{H}(x,u)$ (with 
$I_{3}=-1/2,m=1)$ in the unitary gauge is given by

\begin{align}
L_{H}=& \partial _{\mu }\Phi _{H}^{\dagger }\partial ^{\mu }\Phi _{H}+\mid
\Phi _{H}\mid ^{2}[\frac{g_{1}^{2}}{2}W_{\mu }^{+}W^{\mu -}+  \notag \\
& +g_{2}^{2}q_{B}B_{\mu }^{+}B^{\mu -}+E_{\mu }^{c}E^{c\mu }\mid
Q_{E^{c}}\mid ^{2}  \notag \\
& +\mid Q_{Z}\mid ^{2}Z_{\mu }Z^{\mu }]+V(\Phi _{H}),  \label{63}
\end{align}%
with $Q_{Z}=-\frac{1}{2}\cos \theta _{D}g+g_{3}\varkappa _{H}\sin \theta
_{D},Q_{E^{c}}=g\sin \theta _{D}/2+g_{3}\varkappa _{H}\cos \theta
_{D},q_{B}=4\varkappa _{H}(n+1).$ The interaction of the SM Higgs with the $%
W $ \ bosons remains the same as in the SM. The $E^{c}$ and $B^{\pm }$ gauge
bosons are not decoupled from the SM particles, according to (62) there is a
coupling of the new $E^{c}$ and $B_{\mu }^{\pm }$ bosons to the SM Higgs. In
(62) the SM Higgs self-interaction potential $V(\Phi _{H})=-\mu ^{2}$ $\mid
\Phi _{H}\mid ^{2}+\lambda \mid \Phi _{H}\mid ^{4}$ is included.

We denote the left- handed lepton family with $E_{L}=(e_{L},\mu _{L},\tau
_{L})$ and the right-handed family with $E_{R}=(e_{R},\mu _{R},\tau _{R}).$
The SM Yukawa interaction $L_{Yuk}$ term is given by 
\begin{equation}
L_{Yuk}=-\Sigma _{n_{1}n_{2}}\sigma _{n_{1}n_{2}}\Psi _{M_{E_{L}}}^{\dagger
}\Phi _{M_{H}}\Psi _{N_{_{E_{R}}}}+hc
\end{equation}%
with $\sigma _{n_{1}n_{2}}$ as constant coupling coefficient and with the
IQNs $M_{E_{L}}=\{n_{1},m=-1,j=j_{3}=\frac{1}{2},\varkappa _{E_{L}}\},$ $%
N_{E_{R}}=\{n_{2},m=-2,j=0,\varkappa _{E_{R}}\}$ and with the IQN of the
Higgs: $M_{H}=\{n_{H},m_{H}=1,j_{3}=-\frac{1}{2}.\varkappa _{H}\}$. An
assumption concerning the IQN $\varkappa $ of leptons and the Higgs
particles will be discussed later.

The self-interaction term $V(\Phi _{S})$ and $L_{Yuk}$ do not arise in the
TB tree level approximation but are included by phenomenological reasons in
the same way as in the SM. A microscopic foundation of these
phenomenological terms is an unsolved problem in the SM as well as in the
here presented approach.

\bigskip \textbf{6. Quantization on the tangent bundle}

On the tangent bundle one-particle states of the fermion Dirac field $\Psi
_{f}(x,u)=\sum_{M,s}(\psi _{M,s}(x)\chi _{M}(u)+\psi _{M,s}^{\dagger
}(x)\chi _{M}^{\ast }(u))$ labeled by the three-momentum $\mathbf{p}$ are
described by

\begin{eqnarray}
\Psi _{f}(x,u) &=&\sum\limits_{K}\frac{1}{\sqrt{2E_{M}^{f}V}}%
[a_{K}^{f}u_{M}^{f}(p)\chi _{M}^{f}(u)\exp (i\mathbf{px})  \notag \\
&&+b_{K}^{\dagger f}v_{M}^{f}(p)\chi _{M}^{f\ast }(u)\exp (-i\mathbf{px})],
\label{6,1}
\end{eqnarray}%
where the index $s$ characterizes the helicity $s=\{L,R\}$ and $K=\{M,%
\mathbf{p},s\}.$ $a_{K}^{f}(t)$ is the annihilation operator for a particle
in the interaction representation and $b_{K}^{\dagger f}$ $(t)$\ the
antiparticle creation operator satisfying the anti commutation rules. $%
E_{M}^{f}$ is the single particle energy and V the volume. $u_{M}^{f}(p)$
and $v_{M}^{f}(p)$ are the plane wave solutions of the Dirac equation for
particles and antiparticles, respectively and the eigenfunctions $\chi
_{M}(u)$ are given in (29) and (36). The general construction of states in
the TB indicates that not only leptons but also scalars and gauge bosons
carry the IQN $M_{A}=\{n,m,\varkappa ,j,j_{3}\}.$ This means scalar fields
and gauge fields can be expanded in the analog form as (64).

The structure of the theory based on the TB geometry suggests an
identification of an elementary particle as a state with specific internal
quantum numbers $M$ and a specific mass analogous as in quantum mechanics of
atoms discrete quantum states with different quantum numbers and energy
levels exist. Therefore we do not fix the particle content from the
beginning but let the existence of "exotic" particles open which do not
appear in the SM and are not observed so far. The potential observation of
such particles depends on its parameters as mass and lifetime, but also on
selection rules as discussed later.

Inserting the expansion (64) into the fermion Laplacian (61) the Hamiltonian
is easy to build with an interaction term with gauge particles $g$. For the
unperturbed fermion Hamiltonian we get

\begin{equation}
H_{0}^{f}=\frac{1}{2}\sum\limits_{P}\epsilon _{\mathbf{p}}[a_{P}^{\dagger
f}a_{P}^{f}-b_{P}^{f}b_{P}^{f\dagger }],  \label{6.2}
\end{equation}%
with $P=\{M,\mathbf{p},s\}$ and with the one-particle energy $\epsilon _{%
\mathbf{p}}=\mid \mathbf{p\mid }.$ For a compact representation we introduce
for the gauge particles the denotation: $g=(g^{0},g^{c})$ with $%
g^{0}=(A,Z,E^{c})$ and $g^{c}=(W^{\pm },B^{\pm })$ and for the gauge
potentials $A_{g\mu }=A_{g\mu }^{+}+A_{g\mu }^{-}$ with $A_{g\mu }^{\pm
}=a_{g\mu }^{\pm }(x)\chi _{M_{g}}^{\pm }(u),\chi _{M_{g}}^{+}=(\chi
_{M_{g}}^{-})^{\ast }.$For the interaction Hamiltonian $H_{I}^{f}$ one gets 
\begin{align}
H_{I}^{f}& =\sum_{PRK,g}(c_{K}^{g^{0}}V_{PRK}^{g^{0}}+c_{K}^{g^{0}\dagger
}V_{PR\widetilde{K}}^{g^{0}\ast }+a_{K}^{g^{c}}V_{PRK}^{g^{c}}+  \label{6.3}
\\
& +b_{K}^{g^{c}\dagger }V_{PR\widetilde{K}}^{g^{c}\ast })a_{P}^{f\dagger
}a_{R}^{f}+(c_{K}^{g}U_{PRK}^{g}+c_{K}^{g\dagger }U_{PR\widetilde{K}}^{g\ast
}+  \notag \\
& +a_{K}^{g^{c}}U_{PRK}^{g^{c}}+b_{K}^{g^{c}\dagger }U_{PR\widetilde{K}%
}^{g^{c}\ast })b_{R}^{f}b_{P}^{f\dagger },  \notag
\end{align}%
with $R=\{N,\mathbf{r},s\},$ $K=\{M_{g},\mathbf{k},\mu \},\widetilde{K}%
=\{M_{g},-\mathbf{k},\mu \}$ and 
\begin{equation}
V_{PRK}^{g}=\frac{\overline{u_{P}^{f}}\sigma _{s}^{\mu }u_{R}^{f}\mathbf{%
\epsilon }_{\mu }(\mathbf{k})I_{MNM_{g}}^{f}\delta _{\mathbf{r,p-k}}}{2\sqrt{%
2E_{\mathbf{p}}^{f}E_{\mathbf{r}}^{f}E_{\mathbf{k}}^{g}V}},  \label{6.7}
\end{equation}%
\begin{equation}
U_{PRK}^{g}=\frac{\overline{v_{P}^{f}}\sigma _{s}^{\mu }v_{R}^{f}\mathbf{%
\epsilon }_{\mu }(\mathbf{k})I_{MNM_{g}}^{f}\delta _{\mathbf{r,p+k}}}{2\sqrt{%
2E_{\mathbf{p}}^{f}E_{\mathbf{r}}^{f}E_{\mathbf{k}}^{g}V}}.  \label{6.8}
\end{equation}%
Here we introduced the matrix elements 
\begin{equation}
I_{M_{g}M_{f}N_{f}}^{f}=\int d\mu \chi _{M_{g}}^{g}(u)\chi _{M_{f}}^{f\ast
}(u)\mathbf{Q}_{g}\chi _{N_{f}}^{f}(u),  \label{6.9}
\end{equation}%
with the integration measure $d\mu (u)$ $=d\mu _{SU(2)}d\mu _{E^{c}},d\mu
_{SU(2)}$ $=(16\pi ^{2})^{-1}\sin \theta d\theta d\psi d\varphi $ and $d\mu
_{E^{c}}=(4\pi ^{2})^{-1}\xi d\xi d\phi d\beta .$ The compact representation
(66) includes all possible interactions with gauge bosons $g=A,Z,W^{\pm
},E^{c}$ and $B^{\pm }$.

The interaction Hamiltonian of scalar particles in the unitary gauge is
described by 
\begin{align}
H_{I}^{S}& =-\sum_{PQKR}(c_{P}^{S\dagger }c_{Q}^{S}+c_{\widetilde{P}}^{S}c_{%
\widetilde{Q}}^{S\dagger })  \label{6.10} \\
\Sigma _{g}& [(c_{K}^{g^{0}\dagger }c_{R}^{g^{0}}+c_{\widetilde{K}%
}^{g^{0}}c_{\widetilde{R}}^{g^{0}\dagger })M_{PQKR}^{g^{0}}+  \notag \\
& +(a_{K}^{g^{c}\dagger }a_{R}^{g^{c}}+b_{\widetilde{K}}^{g^{c}}b_{%
\widetilde{R}}^{g^{c}\dagger })M_{PQKR}^{g^{c}}].  \notag
\end{align}%
Here we introduced the symbols $P=\{M_{S},\mathbf{p}\},\widetilde{P}%
=\{M_{S},-\mathbf{p}\},Q=\{M_{S},\mathbf{q}\},\widetilde{Q}=\{M_{S},-\mathbf{%
q}\},K=\{M_{g},\mathbf{k,\lambda }\},,\widetilde{K}=\{M_{g},-\mathbf{%
k,\lambda }\},$

$R=\{M_{g},\mathbf{r,}\sigma \},\widetilde{R}=\{M_{g},-\mathbf{r,}\sigma \}$
and the matrix elements%
\begin{equation}
M_{PQKR\mathbf{\,}}^{g}=\frac{1}{8}q_{g}\frac{I_{M_{S}M_{g}}^{S}\delta
_{\lambda \sigma }\delta (\mathbf{p}-\mathbf{q}+\mathbf{k}-\mathbf{r})}{V%
\sqrt{E_{\mathbf{p}}^{S}E_{\mathbf{q}}^{S}E_{\mathbf{r}}^{g}E_{\mathbf{k}%
}^{g}}},  \label{6.11}
\end{equation}%
with $q_{A}=Q_{A}^{2},q_{Z}=Q_{Z}^{2},q_{E^{c}}=Q_{E^{c}}^{2},Q_{A}=e(j_{3}+%
\frac{1}{2}m),$ $Q_{Z}=g\cos \theta _{D}(\cos ^{2}\theta _{W}j_{3}-\sin
^{2}\theta _{W}\frac{1}{2}m)+g_{3}\sin \theta _{D}\varkappa
,q_{B}=4\varkappa \lbrack n+\frac{1}{2}(1+\mid m\mid )],$ $%
q_{W}=[j(j+1)-j_{3}^{2}],Q_{E^{c}}=(-g\cos ^{2}\theta _{W}j_{3}+g\sin
^{2}\theta _{W}\frac{1}{2}m)\sin \theta _{D}+g_{3}\cos \theta _{D}\mathbf{%
\varkappa }$ \ and 
\begin{equation}
I_{M_{S}M_{g}}^{S}=\int d\mu (u)\chi _{M_{S}}^{S\ast }(u)\chi
_{M_{S}}^{S}(u)\chi _{M_{g}}^{g\ast }(u)\chi _{M_{g}}^{g}(u).  \label{6.12}
\end{equation}

The dependence of the quantized field operators (64) on the eigenfunctions $%
\chi _{M}(u)$ and on the coordinates $u$ of the tangent vectors is a
specific trait of the here presented approach based on the underlying
geometric structure of the TB. The internal symmetries arise here from the
inherent geometrical symmetries of the TB in an analog way as symmetries in
quantum mechanics originate from spacetime symmetries in a given physical
system. This differ in a principle way from standard QFT, therefore we
denote the here presented theory as Tangent Bundle Quantum Field Theory.

Finally we consider the SM Yukawa interaction $L_{Yuk}$ \ term given by $%
(63).$ Inserting the expansion (64) into (63) one get analog expressions as
in the SM but including matrix elements 
\begin{equation}
I_{M_{H}M_{L}N_{R}}^{Y}=\int d\mu \chi _{M_{H}}^{H}(u)\chi _{M_{L}}^{l\ast
}(u)\chi _{N_{R}}^{l}(u),  \label{6.13}
\end{equation}%
The TB eigenfunctions differ for different $n$ and $\varkappa $ but from
(73) we see that the Yukawa interaction is nonzero only if not only leptons
and quarks but also the Higgs particles carry a nonzero E$^{c}$charge $%
\varkappa _{H}.$ The matrix element \ $I_{M_{H}M_{L}N_{R}}^{Y}$ is non-zero
if the relations for the hypercharges $-m_{E_{L}}+m_{\Phi }+m_{E_{R}}=0$ and
for the E$^{c}$ charges $-\varkappa _{E_{L}}+\varkappa _{\Phi }+\varkappa
_{E_{R}}=0$ are fulfilled. The solution of these equations is not unique.
Since interaction processes favor the lowest magnitude of $\varkappa $ and $m
$ we assume here the special solution $m=\varkappa $ for the above given two
equations. For the left-handed lepton family we obtain $\varkappa
_{E_{L}}=-1,$ for the right handed lepton family $\varkappa _{E_{R}}=-2$ and
for the Higgs family $\varkappa _{H}=1$. An important consequence is that
with $\varkappa _{H}\neq 0$ analogous as lepton families also Higgs families
should exist.

\textbf{7. Lepton families, lepton universality and Higgs interaction beyond
the Standard Model}

\textbf{7.1 Lepton interaction and lepton universality}

SM leptons are distinguished by the IQNs of isospin $j$ and $j_{3}$ and weak
hypercharge $m$. Leptons consist of three families, electrons and electron
neutrinos are members of the first family, muons and muon neutrinos of the
second and taus and tau neutrinos of the third family. Different families
exhibit in the SM identical IQNs and properties in the electroweak
interaction with the exception of its masses. In the TB approach in addition
to isospin $I$ and hypercharge $m$ \ the $E^{c}$-charge $\varkappa $ and the
family quantum number $n$ exist. Here the up to now unexplained fact in the
SM that three lepton families exist differentiated only by its mass find an
explanation by the additional family quantum number $n$ for a non-zero $E^{c}
$-charge $\varkappa $. Besides a larger number of families than three could
exists, but its possible observation depends on the mass of these states
with $n\succ 3$ or other possible physical effects..

The interaction of fermions via gauge potentials is described by (66) with
analog expressions as in the SM but with inclusion of the matrix elements $%
I_{M_{g}M_{f}N_{f}}^{f}$ given in (69) depending on the eigenfunctions of
the Laplacian on the group. From (69) selection rules can be derived for
fermion interactions. If the value of the integral $I_{M_{g}M_{f}N_{f}}^{f}$
is zero the interaction is forbidden. These selection rules arise in a
similar way as the selection rules in atomic systems if the transition
moment integral is vanishing which constrains the possible transitions of a
system from one quantum state to another. From these matrix elements we see
that in the electroweak interaction leptons (with $M_{e}=N_{e\text{ }})$
couple only to photons and Z-bosons with $\varkappa _{g}=0,j_{g}=0,m_{g}=0$
and therefore with $\ h_{M_{g}}(u)\equiv 1$ we find $%
I_{M_{A}M_{l}M_{l}}^{l}=Q$ and $I_{M_{Z}M_{l}M_{l}}^{l}=Q_{Z}.$This means
that family universal electroweak coupling of leptons with photons and
Z-bosons is regained in the here presented approach. For the coupling of the
families of left-handed charged leptons and neutrino%
%TCIMACRO{\U{b4}}%
%BeginExpansion
\'{}%
%EndExpansion
s to the charged $W^{\pm }$-bosons we substitute for $M_{l}$ the IQN of the
lef-handed charged leptons and for $N_{l}$ \ that of the neutrinos. The
operator $\mathbf{Q}_{W^{\pm }}=g_{1}\mathbf{J}^{\pm }$ shifts the iso-spin
component $j_{3}$ in such a way that the matrix element (69) is again
independent on the family number $n$. Lepton flavour universality is one of
the distinctive features of the SM and experiments set stringent limits on
processes that violate this universality. Although now every family is
connected with a different IQN $n$ and different eigenfunctions the coupling
of the photons to the leptons remains independent on the family number $n$.

The same behavior we can find for the interaction with the $E^{c}$-boson
with SM leptons. The matrix element (69)\ with $g=E^{c}$ is non-zero only if
the $E^{c}$-boson carry the IQN $\varkappa
_{E^{c}}=0,j_{E^{c}}=0,m_{E^{c}}=0.$

The mass of leptons and its large difference for electrons, myons and $\tau $%
-leptons as well as the non-zero neutrino masses can not be explained as a
tree-level effect. But the occurrence of a family quantum number $n$ and
different eigenfunctions in the here presented approach could open up a
route towards its physical understanding beyond the tree level in a
non-perturbative treatment. This problem is beyond the present paper.

\textbf{7.2 The interaction of Higgs particles with gauge bosons}

The interaction of scalar particles with gauge bosons is described by (70)
with matrix elements given in (71) and (72). Let us first discuss the weak
interaction of the $Z$ and $W^{\pm }$ bosons with the SM Higgs particle. As
explained above the $E^{c}$ charges $\varkappa $ of the $Z$ and the $W$ \
bosons are necessary zero ($\varkappa _{Z}=0,\varkappa _{W^{\pm }}=0).$
However as described in section 6 the Higgs carry a non-zero E$^{c}$ charge
(here $\varkappa _{H}=1$ is assumed$),$ therefore the coefficients $q_{Z}$
in (71) shows a very small deviation from their value in the SM proportional
to $g_{3}^{2}.$

An interesting feature of the presented approach refer to the interaction of
extra gauge bosons \textbf{\ }$E^{c}$ and $B^{\pm }$ with the SM Higgs
described by (70), (71) and (72). As seen the coupling of the Higgs to $%
E^{c} $ and $B^{\pm }$ bosons is allowed.

Note that the existence of a new vector boson is a common feature of many
extensions of the SM (for a review see \cite{e2}). In particular models with
an extra gauge group $U%
%TCIMACRO{\U{b4}}%
%BeginExpansion
{\acute{}}%
%EndExpansion
(1)$ are studied in large number of papers (see e.g. \cite{e2,e3,e4,e5}).

The possible existence of $E^{c}$ \ and $B^{\pm }$ bosons leads to a
fundamental fifth interaction. The parameter space of $E^{c}$ and $B^{\pm }$
masses and the coupling coefficient $g_{3}$ (or the mixing angle $\theta
_{D})$ are constrained by existing data from experiments, and could be found
in a similar way as for $U%
%TCIMACRO{\U{b4}}%
%BeginExpansion
{\acute{}}%
%EndExpansion
(1$) extended models (see. e.g. \cite{p2,p5,dm1,H1,H2,Z,sa, sa2}). Much of
these data hints to the assumption that the coupling coefficient $g_{3}$ and
the mixing angle $\theta _{D}$ are small: $g_{3}\ll 1,\theta _{D}\ll 1$.
This means that the fifth fundamental interaction mediated by the $E^{c}$%
-boson is much weaker than the SM weak interaction.

\emph{\ }

\bigskip \textbf{\ 8. Dark Matter candidates}

Astrophysical and cosmological observations show that the largest part of
matter in our universe is constituted by unknown non-luminous particles
denoted as Dark Matter (DM) that have a very weak interaction with the
visible sector of the universe. Such particles do not exist in the SM, but
there are many attempts for an extension of the SM with possible DM
candidates such as Weakly Interacting Massive Particles (WIMPs), sterile
neutrinos, the lightest neutralinos in super-symmetric models or axions (see
e.g.\cite{z1,z2}). Recently alternative phenomenological models has been
developed as the Dark Sector Model(see e.g. \cite{dm1,ds2,dm3,dm5,d7,d8}) or
the Higgs portal model \cite{dm4,dm5,H1,H2,sa,sa2}.

One of the most notable feature of the generalization of the SM by the gauge
group $SU(2)\otimes $\textbf{\ }$E^{c}(2)$ is the possibility that Dark
Matter candidates lie within the new gauge sector. In the present approach
for the derivation of the corresponding Lagrangians of DM particles no
additional phenomenological model assumption are requested, but only the
particle content with the choice of appropriate IQNs for the DM is
necessary. An obvious way for the assignment of the IQNs to left- and
right-handed Dark Fermions and Dark Scalars can be made by the choice of
zero hypercharge $(m=0)$ and isospin ($j=0)$ but non-zero $E^{c}$- charge $%
\varkappa \neq 0.$ As a result one can expect that similar as SM leptons DM
fermions and DM scalars are grouped in families with the IQN $n=1,2,3$. In
the Laplacian (61), (62) and (63) we substitute for the lepton and Higgs
wavefunctions $\Psi _{s}^{l}(x,u)\rightarrow $ $\Psi _{s}^{l}(x,u)+\Psi
_{s}^{D}(x,u)$ and $\Phi ^{H}(x,u)\rightarrow $ $\Phi ^{H}(x,u)+\Phi
^{S}(x,u).$According to (29) the eigen functions oft the Laplacian on the
group $SU(2)\otimes E^{c}(2)$ with $j=m=0$ take the form 
\begin{align}
\chi _{M}^{D}& =h_{00\varkappa }(\xi ,\phi )=\sqrt{\frac{\varkappa }{\pi }}%
\exp (i\varkappa \beta )  \label{8.1} \\
& \exp (-\frac{\mid \varkappa \mid \xi ^{2}}{2})L_{n}^{0}(\mid \varkappa
\mid \xi ^{2}).  \notag
\end{align}%
With these extensions the Laplacian (60) now is substituted by $L\rightarrow
L_{SM}+L_{D}$ where $L_{SM}$ describe the SM particles and $%
L_{D}=L_{D}^{f}+L_{D}^{S}+L_{D}^{Yuk}$ includes the Lagrangians of dark
fermions, dark scalars and the dark Yukawa term, respectively. In the
following we discuss these Lagrangians.

\textbf{8.1.} \textbf{Dark vector gauge bosons}

In the present approach new vector bosons $E^{c}$ and $B^{\pm }$ arise
naturally by the geometric TB symmetry described by the group $E^{c}(2).$
These particles can be interpreted as DM vector gauge bosons. The Lagrangian
of the DM gauge bosons is given by (54) where the fields $C_{\mu }^{+}$ and $%
C_{\mu }^{-}$ are related with the $A_{\mu }^{c},Z_{\mu }^{c}$ and the $%
E_{\mu }^{c}$ gauge potentials by the relation (52) and (57). The $E^{c}$
and $B^{\pm }$ gauge bosons are not decoupled from the SM particles,
according to (62) there is a coupling to the SM Higgs, but also the
interaction of the E$^{c}$ boson with leptons with a very small coupling
constant $g_{3}$ is allowed. Note that leptons interact also directly to the
Higgs due to the Yukawa interaction in (63) and therefore via (62)
indirectly couple to the $E^{c}$ and $B_{\mu }^{\pm }$ bosons.

From (40),(50),(54) and (57) we can see that the non-Abelian DM vector
bosons with the gauge potentials $E_{\mu }^{c}$ and $B_{\mu }^{\pm }$
interact with each other but also with the SM gauge bosons $A$, $Z$ and $%
W^{\pm }.$

Note that the hypothesis of self-interacting DM (in contrast to
collisionless cold Dark Matter) enables to resolve a number of conflicts
between observations and predictions of collisionless DM simulations \cite%
{s1,ss1} and has also been assumed as light thermal DM relicts \cite{s2}.

\textbf{8.2.} \textbf{Dark fermions}

We assume that DM fermions with vanishing hyper-charge and isospin ($%
j=0,m=0) $ but nonzero $E^{c}-$charge ($\varkappa \neq 0)$ could exist. The
Lagrangian of the family of DM fermions is given by

\begin{equation}
L_{f}^{D}=i\sum\limits_{Ms}\Psi _{Ms}^{D\dagger }\sigma _{s}^{\mu }[\frac{%
\partial }{\partial x^{\mu }}+ig_{3}\varkappa _{D}(\sin \theta _{D}Z_{\mu
}+\cos \theta _{D}E_{\mu }^{c}\mathbf{)}]\Psi _{Ms}^{D},  \label{8.4}
\end{equation}%
with $M_{D}=(n,0,\varkappa _{D},0,0).$As one can see different types of DM
fermions with non-zero $E^{c}$-charges $\varkappa _{D}$ are predicted. For
every DM fermion with given $E^{c}$-charge $\varkappa _{D}$ a DM fermion
family with $n=1,2,...$ could exist which couple to the SM $Z$ gauge
potential with the coupling coefficient $g_{3}\varkappa _{D}\sin \theta _{D}$
and and to the $E^{c}$gauge potential with the coupling coefficient $%
g_{3}\varkappa _{D}\cos \theta _{D}.$ Analog as the relation (69) one can
derive corresponding selection rules for the interaction of dark fermions
with gauge bosons.

The construction of a gauge and Lorentz invariant mass term for DM fermions
in a renorrmalizable Lagrangian can be done in a similar way as in the SM
using a modified Yukawa interaction term and different IQNs for right and
left-handed DM fermions. Since the SM Higgs carry iso-spin and hypercharge a
DM Yukawa interaction term $L_{Yuk}^{D}$ can not be constructed from the SM
Higgs but instead a scalar DM with vanishing hyper-charge and isospin ($%
j=0,m=0)$ but nonzero $E^{c}-$charge ($\varkappa _{S}\neq 0)$.Therefore the
Yukawa interaction term for scalar DM can be expressed as 
\begin{equation}
L_{Yuk}^{D}=-\Sigma _{n_{1}n_{2}}\sigma _{n_{1}n_{2}}^{D}\Psi
_{M_{D_{L}}}^{\dagger }\Phi _{M_{S}}\Psi _{N_{_{D_{R}}}}+hc  \label{8.6}
\end{equation}%
where the sum is over the DM fermion family members with identical $%
\varkappa $, $\sigma _{n_{1}n_{2}}^{D}$ are constant coupling coefficients.
For the vacuum IQNs of the dark scalar we assume the IQNs $%
M_{S}=\{n_{S}=0,m_{S}=0,j_{3}=0.\varkappa _{S}=1\}.$ This suggest the
following assigned of the IQNs for the family of left-handed and
right-handed dark fermions: $M_{D_{L}}=\{n_{1},m=0,j=j_{3}=0,\varkappa
_{D_{L}}=-1\},$ $N_{D_{R}}=\{n_{2},m=0,j=0,\varkappa _{D_{R}}=-2\}$.

\textbf{8.3.} \textbf{Dark scalars}

The Lagrangian of DM scalars with $j=0,m=0$ but nonzero $\varkappa \neq 0$
and the family number $n$ is given by%
\begin{align}
L_{S}^{D}& =\partial _{\mu }\Phi _{S}^{D\dagger }\partial ^{\mu }\Phi
_{S}^{D}+\mid \Phi _{S}^{D}\mid ^{2}[g_{2}^{2}4\varkappa (n+\frac{1}{2}%
)B_{\mu }^{+}B^{\mu -}  \label{8.3} \\
& +Z_{\mu }Z^{\mu }(g_{3}\varkappa \sin \theta _{D})^{2}\mathbf{+}E_{\mu
}^{c}E^{c\mu }((g_{3}\varkappa \cos \theta _{D})^{2}]  \notag \\
& +V(\Phi _{S}^{D},\Phi _{H}),  \notag
\end{align}%
where $V(\Phi _{S}^{D},\Phi _{H})$ is the the nonlinear Higgs-type potential
for the DM scalar including a possible coupling of the SM Higgs to the DM
scalar: 
\begin{equation}
V(\Phi _{S}^{D},\Phi _{H})=-\mu _{S}^{2}\mid \Phi _{S}\mid ^{2}+\lambda
_{S}\mid \Phi _{S}\mid ^{4}+\lambda _{SH}\mid \Phi _{S}\mid ^{2}\mid \Phi
_{H}\mid ^{2}  \label{8.4a}
\end{equation}%
As seen in (77) coupling of DM scalars to DM gauge vector bosons $E^{c}$ and 
$B^{\pm }$ is allowed. But with a very small coefficient $g_{3}\sin \theta
_{D}$ there exists also a coupling to the SM $Z$ boson arising from the
diagonalization of the Laplacian for the Higgs. The coupling of a DM scalar
particle to gauge bosons $g=(E^{c},Z,B^{\pm })$ is described by (70), (71)
and (72) and by the eigenfunctions $\chi _{M}^{D}(u)$ as given in (74).

In order to generate the gauge boson and DM fermion mass the potential for
the scalar DM should \ develop a nonzero VEV. Taking the extreme of $V(\Phi
_{S}^{D},\Phi _{H})+V(\Phi _{H})$ a non-zero VEV $\prec \Phi _{H}\succ =\Phi
_{H}^{0}$ $\ $and $\prec \Phi _{S}\succ =\Phi _{S}^{0}$ can be calculated
which are given by 
\begin{eqnarray}
(\Phi _{H}^{0})^{2} &=&\frac{4\mu _{H}^{2}\lambda _{S}-2\mu _{S}^{2}\lambda
_{SH}}{4\lambda _{S}\lambda _{H}-\lambda _{SH}^{2}},  \label{8.5a} \\
(\Phi _{S}^{0})^{2} &=&\frac{4\mu _{S}^{2}\lambda _{H}-2\mu _{H}^{2}\lambda
_{SH}}{4\lambda _{S}\lambda _{H}-\lambda _{SH}^{2}}.  \notag
\end{eqnarray}%
Choosing the unitary gauge and expanding the Higgs and the DM scalar around
their VEVs by $\Phi _{H}=\Phi _{H}^{0}+\Phi _{h},$ $\Phi _{S}=\Phi
_{S}^{0}+\Phi _{s}$ the mass squared matrix for the SM Higgs and for the DM
scalars is given by%
\begin{equation}
M^{2}=%
\begin{bmatrix}
2\lambda _{H}(\Phi _{H}^{0})^{2} & \lambda _{SH}\Phi _{H}^{0}\Phi _{S}^{0}
\\ 
\lambda _{SH}\Phi _{H}^{0}\Phi _{S}^{0} & 2\lambda _{S}(\Phi _{S}^{0})^{2}%
\end{bmatrix}
\label{8.5c}
\end{equation}

This matrix can be diagonalized by

\begin{eqnarray}
\Phi 
%TCIMACRO{\U{b4}}%
%BeginExpansion
{\acute{}}%
%EndExpansion
_{h} &=&\cos \beta \Phi _{h}+\sin \beta \Phi _{s}  \label{8.5d} \\
\Phi 
%TCIMACRO{\U{b4}}%
%BeginExpansion
{\acute{}}%
%EndExpansion
_{s} &=&-\sin \beta \Phi _{h}+\cos \beta \Phi _{s}  \notag
\end{eqnarray}%
With the mixing angle 
\begin{equation}
\tan \beta =\frac{\lambda _{SH}\Phi _{H}^{0}\Phi _{S}^{0}}{\lambda _{S}(\Phi
_{S}^{0})^{2}-\lambda _{H}(\Phi _{H}^{0})^{2}+\Lambda \ \ }  \label{8.5f}
\end{equation}%
with $\Lambda =\sqrt{(\lambda _{S}(\Phi _{S}^{0})^{2}-\lambda _{H}(\Phi
_{H}^{0})^{2})^{2}+(\lambda _{SH}\Phi _{H}^{0}\Phi _{S}^{0})^{2}}.$ The
masses for the diagonalized mass eigenstates are 
\begin{eqnarray}
M_{h%
%TCIMACRO{\U{b4}}%
%BeginExpansion
{\acute{}}%
%EndExpansion
} &=&\lambda _{S}(\Phi _{S}^{0})^{2}+\lambda _{H}(\Phi _{H}^{0})^{2}+\Lambda
\   \label{8.5g} \\
M_{s%
%TCIMACRO{\U{b4}}%
%BeginExpansion
{\acute{}}%
%EndExpansion
} &=&\lambda _{S}(\Phi _{S}^{0})^{2}+\lambda _{H}(\Phi _{H}^{0})^{2}-\Delta
\   \notag
\end{eqnarray}%
$.$

Spontaneous symmetry breaking of the SM Higgs field and the DM scalar leads
to the generation of masses for the gauge bosons $Z$, $W^{\pm }$ as well as
for the dark vector bosons $E^{c}$ and $B^{\pm }$ which contain
contributions from the Higgs VEV as well as from the DM scalar VEV%
\begin{equation}
M_{W^{\pm }}^{2}=(\Phi _{H}^{0})^{2}g_{1}^{2}/2  \label{9.1}
\end{equation}

\begin{eqnarray}
M_{Z}^{2} &=&2(\Phi _{H}^{0})^{2}[\frac{1}{2}g\cos \theta _{D}-g_{3}\sin
\theta _{D}]^{2}+  \label{9.2} \\
&&+2(g_{3}\varkappa \sin \theta _{D})^{2}(\Phi _{S}^{0})^{2}  \notag
\end{eqnarray}%
\begin{eqnarray}
M_{E^{c}}^{2} &=&2(\Phi _{H}^{0})^{2}[\frac{1}{2}g\sin \theta _{D}+g_{3}\cos
\theta _{D}]^{2}+  \label{9.3} \\
&&+2(g_{3}\varkappa \cos \theta _{D})^{2}(\Phi _{S}^{0})^{2}  \notag
\end{eqnarray}%
\begin{equation}
M_{B^{\pm }}^{2}=2g_{2}^{2}[2(\Phi _{H}^{0})^{2}+(\Phi _{S}^{0})^{2}]
\label{9.4}
\end{equation}

The mass of the $W^{\pm }$ bosons is identical with its value in the SM. For
the Z boson we obtain from (85) a mass with a small deviation from the SM
proportional to ($g_{3})^{2}$ for $g_{3}\ll 1:$ $\delta M_{Z}=(\Phi
_{H}^{0})4g_{3}^{2}/\sqrt{2}g$ which can be used for the determination of
experimental bounds of the coupling coefficient $g_{3}.$ As seen in (87)
with $\varkappa _{H}=1,m_{H}=1,n_{H}=0,$ the $B^{\pm }$ boson gets a mass
proportional to the SM coupling constant $g_{2}.$ Without the VEV of the
dark scalar (with $\Phi _{S}^{0}\rightarrow 0)$ one gets $M_{B^{\pm }}=\Phi
_{H}^{0}g_{2}2\simeq 116.24$GeV (using $\Phi _{H}^{0}=180GeV)$. According
(86) the mass of the E$^{c}$ boson given by $M_{E^{c}}\simeq 2\Phi
_{S}^{0}g_{3}$\emph{\ }for $g_{3}\ll 1$ is proportional to $g_{3}.\ $

Albeit on a different theoretical basis, the described predictions show some
similar features that arise in various scenarios for DM physics denoted as
the Dark Sector (see e.g. \cite{dm1,ds2,dm3,dm5,d7,d8}), the vector Higgs
portal (see e.g. \cite{dm4,dm5,H1,H2,sa,sa2} and the Z portal (see e.g. \cite%
{Z,sa}\textit{). }The Dark Sector hypothesis assumes that DM interact only
through a new $U_{D}(1)$ force with a hypothetical "Dark photon" as gauge
boson but SM matter do not interact directly with DM particles but can
interact indirectly via a kinetic mixing term in the Lagrangian \cite{ds2}$.$
In the Higgs portal model a DM massive vector boson associated with a hidden
U%
%TCIMACRO{\U{b4}}%
%BeginExpansion
\'{}%
%EndExpansion
(1) symmetry couple to the SM Higgs and in the Z-portal dark matter model a
DM fermion interact directly with the SM Z-boson. Ref. \cite{sa,sa2} \textit{%
\ }reports that the Higgs portal model is compatible with the available
data. Besides in \cite{sa} also acceptable regions of parameters for the
Z-portal model for the interaction of DM fermions with the SM Z boson were
reported. From these results similar conclusions can be drawn for the
acceptable parameter region of the Lagrangians (75) and (77).

A comparison of the here presented approach with the Higgs portal and the Z
portal models shows some common properties but also clearly distinct
features and principal differences. Whereas the majority of DM models can be
considered as a minimal extension of the SM based on a phenomenological
model assumption to understand the mechanism of annihilation, scattering and
possible decays of SM particles the idea of the present approach is to
describe fundamental interactions of SM particles as well as DM particles in
a uniform way without phenomenological model assumption within the
geometrical structure of the TB with the same symmetry group $SU(2)\otimes
E^{c}(2)$. As seen in the Lagrangian (62) the SM Higgs interact with the
dark vector bosons $E^{c}$ and $B^{\pm }$ and in the Lagrangian (77) a DM
scalar and in (75) a DM fermion with the SM $Z$ boson. The extra DM vector
bosons $E^{c}$ and $B^{\pm }$ are nonabelian gauge bosons and interact to
the SM gauge bosons. Due to the existence of families the SM leptons carry
an $E^{c}$ charge. A nonzero Yukawa interaction term hints that also the
Higgs carry a nonzero $E^{c}$ charge $\varkappa $ and therefore a family of
Higgs particles could exist. The coupling of the Higgs and DM scalars in
(62) and (77) with the $E^{c}$ and $B^{\pm }$ arise in an intrinsic way by
the symmetry group while interactions of the Z-Boson with the DM $E^{c}$ and 
$B^{\pm }$ boson or a scalar DM particle arise due to the diagonalization of
the Higgs Lagrangian by the transformations (57).

Above we discussed only DM candidates with zero isospin and hypercharge, but
there exists the possibility that DM particles basically interact by the
weak coupling to SM particles, this means they are electrically neutral but
carry hypercharge and isospin satisfying the relation $Q_{A}=e(j_{3}+\frac{1%
}{2}m)=0.$ This includes right handed sterile neutrinos with $j_{3}=\frac{1}{%
2}$ and $m=-1.$

\textbf{9.} \textbf{Conclusions}

The present paper is based on the hypothesis that the tangent bundle is the
underlying geometrical structure for the description of the fundamental
physical interactions. The internal (gauge) symmetries are not inserted as
an extra theory constituent given externally a priory by phenomenological
reasons like in the SM but come out from the inherent geometrical structure
of the TB with symmetries described by the group $SO(3,1)$. Projective
irreducible representations of this group can be constructed by using the
little groups $SU(2),E^{c}(2)$ and $SU(1,1)$. Using the covariant derivative
given by the operators on the transformation group $G=SU(2)\otimes E^{c}(2)$%
, and the corresponding connection coefficients (gauge potentials) a
generalized theory of the electroweak interaction is derived. Since the SM
arises (without phenomenological assumptions) as a limiting case the
presented approach answer the question why the gauge group of electroweak
interaction in the SM is $G=SU(2)\otimes U(1)$.

In the TB approach wave functions depend on the space-time coordinates $x$
as well as on the coordinates of the tangent fibers $u$. The known SM $Z$
and $W^{\pm }$ gauge bosons can be found again but in addition new extra
gauge bosons $E^{c}$ and $B^{\pm }$ are predicted which constitute a fifth
fundamental interaction. In addition to the SM quantum numbers of isospin
and hypercharge there exist the $E^{c}$-charge $\varkappa $ and the family
quantum number $n$. The existence of the family IQN $n$ in the TB approach
shed light on the mysterious appearance of lepton families in the SM
requiring a distinct IQN for every family. However the large mass difference
of different families cannot be explained as a tree-level effect but
requires a non-perturbative quantum loop treatment which is beyond the
present paper. The existence of families requires that leptons carry a
non-zero $E^{c}$ -charge $\varkappa _{l}$. A selection rule from the Yukawa
interaction indicate that also the Higgs particle carry the $E^{c}$ -charge $%
\varkappa _{H}=1$ and therefore in the present approach a family of Higgs
particles is predicted. In contrast, SM and $E^{c}$ gauge bosons carry zero $%
E^{c}$-charge $\varkappa _{g}=0$. The derived selection rule reveals that
family universal coupling of leptons persists for the interaction with $SM$
gauge bosons.

An important prediction of the theory presented is the possibility of
identifying candidate stable or unstable hypothetical DM fermions and DM
scalars with zero hypercharge and zero isospin but non-zero $E^{c}$ -charge $%
\varkappa \neq 0$ which should necessarily grouped in different families
with different family numbers $n=1,2,...$The new non-Abelian vector bosons $%
E^{c}$ and $B^{\pm }$ can be interpreted as DM vector gauge bosons. These
hypothetical bosons are not decoupled from the SM particles, but there is a
weak coupling to the SM Higgs and to the SM Z boson. Besides also the
coupling of leptons with the $E^{c}$ bosons with a small coupling
coefficient $g_{3}$ is allowed . DM vector bosons interact with each other
but also with the SM gauge bosons. Spontaneous symmetry breaking of the SM
Higgs and the DM scalar predicts not only the masses of the $Z$ and $W^{\pm
} $ bosons but also that of the dark gauge bosons $E^{c}$ and $B^{\pm }.$The
precondition of nonzero $E^{c}$ charges $\varkappa \neq 0$ of DM fermions
and DM scalars indicates that analogical as lepton families also DM fermion
and DM scalar families should exist.

Finally, the approach presented is linked with the geometrization program of
physics based on a single hypothetical principle that the tangent bundle
with the symmetry group $SO(3,1)\rtimes T(3,1)$ is the fundamental
geometrical structure for an unified description of all fundamental physical
interactions. On the one hand as briefly explained in section 3 the tangent
bundle is the geometrical fundament for teleparallel gravity gauge theory
based on translational transformations $T(3,1)$ of tangent vectors along the
fiber axis \cite{t1,g5,t2,t3,M} which is fully equivalent to the Einstein
gravity theory. On the other hand here a generalized theory of electroweak
interaction and dark matter is presented based on the little groups of $%
SO(3,1)$. Therefore gravity, electroweak interaction and Dark Matter are
described by the same fundamental geometrical structure of the TB. Note that
strong interaction with the gauge group $SU(3)$ is in this frame still
missing. The color group $SU(3)$ of Quantum Chromodynamics can not be
described as a geometrical symmetry in the TB in a way as the $SU(2)\otimes
E^{c}(2)$ group leaving the scalar product (3) invariant. However the $SU(3)$
symmetry could be hidden in the fundamentals of the Tangent Bundle geometry
in a surprising way arising as an emergent symmetry similar as Chern-Simon
gauge fields originate by the anomalous Quantum Hall Effect in Solid State
Theory (see e.g. \cite{QH1,QH3,QH4,QH5}.The key for understanding of this
assumption is the fact that the eigenfunction of the E$^{c}(2)$ group given
in (29) have the same form as the solution of the 2D Schr\"{o}diner equation
for electrons in a perpendicular external magnetic field. The solution (29)
describes the vertical subspace for a single tangent fiber at a fixed
spacetime point, but if we combine all tangent fibers at all spacetime
points we get an equation with the analog form as the multi-particle Schr%
\"{o}dinger equation of a 2D quantum Hall system and with the account of the
three iso-spin components of fermions that of a three-layer Quantum Hall
system \cite{QH1}. This explains the astonishing analogy of fractional
charge quantization of quarks with the anomalous Qantum Hall effect \cite%
{QH5}. In this approach emergent  effective gauge fields \cite%
{CS1,CS2,QH1,QH3} (arising here in the vertical subspace of tangent vectors
in the TB) originate in an intrinsic way and bound the bare quarks to two
vortices constituting composite quarks. These gauge fields and the SU(3)
symmetry is assigned to the base spacetime of the TB via the density
distributions and could appear as the local SU(3) color symmetry of Quantum
Chromodynamics.


\begin{thebibliography}{99}
\bibitem{r1} E. Lubkin, Ann.Phys. (N.Y.)\textbf{\ 23}, 233 (1963),

\bibitem{r2} A.Trautman, Rep. Math. Phys.\textbf{\ 1,} 29 (1970),

\bibitem{r3} W. Drechsler W and M. E. Meyer; "Fibre Bundle Techniques in
Gauge Theories", Lecture Notes in Physics, Springer1977,

\bibitem{r5} M. Daniel and C. M. Viallet, Rev. Mod. Phys. \textbf{52}, 175
(1980),

\bibitem{r6} M. Nakahara, "Geometry, Topology and Physics", Institute of
Physics , Bristo 1990,

\bibitem{r7} Bo-Yuan Hou, "Differential Geometry for Physicists", Singapore
1997,

\bibitem{Col} S.Coleman and J.Mandula, Phys. Rev.189, 1251 (1967),

\bibitem{Ko} S. Kobayashi an K. Nomizu, "Foundation of Differential
Geometry", Interscience Publishers (1963),

\bibitem{g4} Chris J Isham, "Modern Differential Geometry for Physicists",
World Scientific (1999),

\bibitem{Sch} J.A. Schouten, "Ricci Calculus", Springer Verlag, Berlin
(1954),

\bibitem{t1} K. Hayashi and T. Nakno, Prog Theor. Phys. \textbf{38}, 491
(1967),

\bibitem{t2} Y. M. Cho, Phys. Rev.D \textbf{14}, 2521 (1976),

\bibitem{g5} T. Dass, Pramana \textbf{23}, 433 (1984),

\bibitem{t3} R. Aldrovandi, J.G. Pereira "Teleparallel Gravity; an
Introduction" (Springer, Dordrecht, 2012),

\bibitem{M} J. W. Maluf, Annalen Phys. \textbf{525,} 339 (2013),
arXiv:1303.3897 [gr-qc],

\bibitem{g2} T. W. Kibble, J. Math. Phys.\textbf{\ 2}, 212, (1961),

\bibitem{g3} F. W. Hehl, J.D. McCrea, E. W. Mielke and Y. Ne%
%TCIMACRO{\U{b4}}%
%BeginExpansion
\'{}%
%EndExpansion
eman, Phys. Rep. \textbf{258}, 1 (1995),

\bibitem{g6} F. W. Hehl, J.D. McCrea, E. W. Mielke and Y. Ne%
%TCIMACRO{\U{b4}}%
%BeginExpansion
\'{}%
%EndExpansion
eman, Phys.Rev. D \textbf{48}, 48 (1993),

\bibitem{r8} E. P. Wigner, Ann. Math. \textbf{40}, 149 (1939),

\bibitem{r9} V. Bargman, Ann. Math. \textbf{59}, 1 (1954),

\bibitem{r10} H. Hoogland, J. Phys. A Math. Gen. \textbf{11}, 1557 (1978),

\bibitem{r11} J. F. Carina, M. A. del Olmo and M Santander, J.Phys.A:
Math.Gen. \textbf{17}, 3091 (1984),

\bibitem{Q1} J. J. Quinn, A. W\'{o}js, K.-S. Yi, and G. Simion, Physics
Reports \textbf{481},29 (2009),

\bibitem{h2} V.Aldaya, J. Navarro-Salas, J. Bisquert and R. Loll, J. Math.
Phys. \textbf{33}, 3087 (1992),

\bibitem{f1} I.M. Shapiro, R.A. Minlos and Z.Ya Shapiro, "Representations of
the Rotation and Lorentz Groups and their Applications", Pergamon Press 1963,

\bibitem{f2} W. J. Holman and L. C. Biedenhan, Jr. Annals of Physics \textbf{%
39}, 1, (1966),

\bibitem{IP1} D. Cangemi and R. Jackiw, Phys.Rev.Lett.69 (1992) 233,

\bibitem{IP2} C. Nappi and E. Witten, Phys. Lett. B293 (1992) 309,

\bibitem{IP3} A. Tseytlin, Nuclear Phys. B450, 231 (1995),

\bibitem{IP4} K. Sfetsos, Phys. Lett. B324 (1994) 335,

\bibitem{IP5} P. A. Grassi,Nucl. Phys. B560, 499 (1999),

\bibitem{IP6} F. Ferrari, J. Math. Phys. 44,138 (2003),

\bibitem{IP7} F. R. Ruiz, Eur. Phys. J. C (2016) 76:94,

\bibitem{p2} R. Essig et al., JHEP \textbf{02}, 009 (2011),

\bibitem{p5} J. D. Bjorken et al. Phys. Rev. D \textbf{80}, 07501 (2009),

\bibitem{dm1} J. Alexander et al.: Dark Sectors 2016 Workshop: Community
Report, ArXiv:1608.08632v1,

\bibitem{e2} P. Langacker, Rev. Mod. Physics \textbf{81}, 1199 (2009),

\bibitem{e3} T. Appelquist, B.A. Dobrescu and A. R. Hopper, Phys. Rev. D 
\textbf{68}, 035012 (2003),\textbf{\ }

\bibitem{e4} K. S. Babu, C. Kolda and J. March-Russell, Phys. Rev. D \textbf{%
57}, 6788 (1998),

\bibitem{e5} F. Blas and M Perez-Victoria F. del Aguila, J. de Blas and M
Perez-Victoria JHEP \textbf{1009}, 033 (2010)),

\bibitem{z1} G. Bertone, D. Hooper and J. Silk, Particle dark matter:
evidence, candidates and constraints. Phys. Rep. \textbf{405} (2005) 279,

\bibitem{z2} Stefan Profumo, "An Introduction to Particle Dark Matter",
World Scientific (2017),

\bibitem{s1} D.N.Steinhardt, J.Paul, Phys. Rev. Lett. \textbf{84}, 3760
(2000),

\bibitem{ss1} S. Tulin and H.-B. Yu, Phys. Rep. \textbf{730}, 1 (2018),

\bibitem{s2} Y. Hochberg et al., Phys. Rev. Lett. \textbf{113}, 171301
(2014),

\bibitem{dm4} V. Silveira and A. Zee, Phys. Lett. B \textbf{161}, 136 (1985),

\bibitem{dm5} Thomas Hambye, JHEP \textbf{01} (2009) 028,

\bibitem{ds2} B. Holdom, Phys. Lett. B \textbf{166}, 196 (1986),

\bibitem{dm3} P-F Yin and S.-H. Zhu, Front. Phys. \textbf{11}, 111403 (2016),

\bibitem{d7} N. Bernal et al., JCAP \textbf{1603}, 028 (2016),

\bibitem{d8} S.-M. Choi et al., JHEP \textbf{1710} (2017) 162,

\bibitem{H1} S. Kanemura et al., Phys. Rev. D \textbf{82}, 055026 (2010)

\bibitem{H2} O.Lebedev, H. M:Lee and Y.Mambrini Phys.Lett.B \textbf{702},
570 (2012)

\bibitem{Z} G.Arcadi, Y. Mambrini and F. Richard, arXiv:1411.2985v1

\bibitem{sa} J.Ellis, A. Fowlie, L.Marzola and M.Raidal, Phys. Rev.D \textbf{%
97}, 115014 (2018)

\bibitem{sa2} P.Athron et al.Eur.Phys.J. C \textbf{79}, 38 (2019),

\bibitem{QH5} R. B.Laughlin, Phys. Rev.Lett.\textbf{50}.1395 (1983)

\bibitem{QH1} Z. F. Ezawa, "Quantum Hall Effects", World Scientific, 2013

\bibitem{QH3} O.Heinonen (ed.), "Composite Fermions", World Scientific, 1998

\bibitem{QH4} J. K. Jain, "Composite Fermions", Cambridge University Press,
2009

\bibitem{CS1} S.C. Zhang, H.Hansson and S.Kivelson, Phys. Rev.Lett.\textbf{62%
}, 82, (1989)

\bibitem{CS2} A. Lopez and E. Fradkin, Phys. Rev.B \textbf{44}, 5246 (1995)
\end{thebibliography}
\end{document}